\documentclass{aa}
\usepackage{epsfig}
\epsfclipon
\def\kms{km~s$^{-1}$}
\def\cm2{cm$^{-2}$}
\def\hkpc{$~h_{50}^{-1}$ kpc}
\def\hs{HS 1216+5032}

\begin{document}

   \thesaurus{12
              (12.03.3)    
               11
               11.17.3;  
               11.17.1;  
               11.17.4)} 

   \title{HST spectroscopy of the double QSO HS 1216+5032
AB\thanks{Based on 
observations with the NASA/ESA Hubble Space Telescope, obtained at the
STScI, which is operated by AURA, Inc., under NASA contract
NAS5--26\,555.}}

   \author{Sebastian Lopez\thanks{Present address: Departamento de
          Astronom\'{\i}a, Universidad de Chile, Casilla 36-D,
          Santiago, Chile } 
          \and
          Hans-J\"urgen Hagen
          \and
          Dieter Reimers
          }

   \offprints{S. Lopez, e-mail: slopez@das.uchile.cl}

   \institute{
              Hamburger Sternwarte, Universit\"at Hamburg, Gojenbergsweg 112, 
              21029 Hamburg, Germany 
             }

   \date{Received; accepted}

\titlerunning{HST spectroscopy of HS 1216+5032 AB}
   \maketitle

   \begin{abstract}
We report on {\it Hubble Space Telescope} Faint Object Spectrograph
observations of the double QSO  HS 1216+5032 AB ($z_e({\rm
A})=1.455$; $z_e({\rm B})=1.451$, and angular separation
$\theta=9\farcs1$). The spectral coverage is 910 \AA{} to 1340 \AA{}
in the QSO rest-frame. An unusual broad-absorption-line (BAL) system is
observed only in the B component: maximum outflow velocity of $\sim5\,000$
\kms; probably a mixture of broad and narrow components. Observed ions
are: \ion{H}{i}, \ion{C}{ii}, \ion{C}{iii}, \ion{N}{iii}, \ion{N}{v}, 
\ion{O}{vi}, and possibly \ion{S}{iv} and \ion{S}{vi}. We also discuss
two outstanding 
intervening systems: (1) a complex \ion{C}{iv}
system at $z=0.72$ of similar strength in A and B, with a velocity  span of
$1500$ \kms{} along the lines of sight (LOSs; LOS separation: 
$S_{\bot}[z=0.72]\approx 75~h_{50}^{-1}$ kpc); and (2) a possible
strong \ion{Mg}{ii} system at $z=0.04$ observed in B only, presumably
arising in a damped Ly$\alpha$ system.

We assume \hs{} is a binary QSO but discuss the possibility of a
gravitational lens system. 
The size of Ly$\alpha$ forest clouds is constrained using
$S_{\bot}\approx 80~h_{50}^{-1}$ kpc  
at redshifts between $z=1.15$ and
$1.45$. Four Ly$\alpha$ systems not associated with metal lines and
producing lines  with $W_0>0.17$ \AA{} are observed in
both spectra, while five appear in only one spectrum. This sample,
although scarce due to the redshift path blocked out by the BALs in B,
allows us to place upper limits on the transverse cloud sizes. 
Modelling the absorbers as non-evolving spheres, a maximum-likelihood
analysis yields a most probable cloud diameter 
$D=256~h_{50}^{-1}$ kpc and $2 \sigma$ bounds of
$172<D<896~h_{50}^{-1}$ kpc. If the
clouds are modelled as filamentary structures, the same analysis
yields lower transverse
dimensions by a factor of two. Independently of the maximum-likelihood
approach, the equivalent width differences provide evidence for
coherent structures. 
The suggestion that the size
of Ly$\alpha$ forest clouds increases with decreasing redshift is not
confirmed. Finally, we discuss two $z_a\approx z_e$ Ly$\alpha$ systems
observed in both QSO 
spectra.

   \end{abstract}

\section{Introduction}
\label{sec_int}

HS 1216+5032 was discovered in the course of the Hamburg Quasar Survey
(Hagen et al.~\cite{Hagen1}). It was later confirmed as a double QSO at
$z_e=1.45$ by Hagen et al. (\cite{Hagen}) through direct images and
low-resolution (FWHM $\approx20$ \AA) spectroscopy in the optical range. The
$B$ magnitudes of the 
bright and faint QSO images (hereafter ``A'' and ``B'', respectively)
are $B_{\rm A}=17.2$ and $B_{\rm B}=19.0$. The separation angle
between A and B is $\theta=9\farcs1$, and the emission redshifts are
$z_e({\rm A})=1.455$ and $z_e({\rm B})=1.451$. From the available optical data
of HS 1216+5032 AB there appears to be 
some evidence favoring its physical-pair nature instead of a
gravitational lens origin (this subject 
will be discussed in \S\ref{1216_lensed?}), so the projected
distance\footnote{Throughout this paper we assume an
$(\Omega_0,\Lambda_0)=(1,0)$ cosmology and define $h_{50}\equiv
H_0/(50$~\kms Mpc$^{-1})$.} between LOSs as a function 
of redshift is known ($78~h_{50}^{-1}$ kpc at $z_e$).

This paper presents ultraviolet (UV) spectra of QSO A and B taken with
the Faint Object Spectrograph (FOS) on-board the {\it Hubble Space
Telescope} ({\it HST}). The prime goal of these observations was 
originally to investigate the Ly$\alpha$ forest in the 
little explored redshift range between $z=0.8$ and $z=1.4$. 
In particular, the projected separation between the 
LOSs to HS 1216+5032 A and B is convenient because it samples well the
expected Ly$\alpha$ cloud sizes of hundreds of kpc. 

For the redshift range accessible from the ground, 
a pioneering work on cloud sizes was made by Smette et
al. (\cite{Smette1}) using the gravitationally lensed double QSO UM
673 ($\theta=2\farcs2$) with known lens geometry. All observed
Ly$\alpha$ lines were observed to be 
common to both spectra and a statistical lower limit of $D>12$
$h_{50}^{-1}$ kpc could be set for the transverse
sizes (spherical clouds, no evolution). A similar result (Smette et
al.~\cite{Smette}) was  
found for HE 1104$-$1805 AB ($\theta=3\arcsec$) but here the position
of the lens $z_{\rm l}$ was 
unknown. A transverse size of $D>60~h_{50}^{-1}$ kpc was estimated for
$z_{\rm l}\approx 1$.  

Surprisingly, QSO pairs at very large projected
distances seem to still show 
Ly$\alpha$ lines common to both spectra. In the spectra of LB 9605 and
LB 9612 ($\theta=1\farcm65$), Dinshaw et
al. (\cite{Dinshaw1}) find 5 such lines within $400$ \kms{} and derive
a most probable diameter of
$1520$ $h_{50}^{-1}$ kpc at $1<z<1.7$. If the lines arise in different
absorbers, however, an upper limit of $\sim1140$
$h_{50}^{-1}$ kpc is derived. Further studies 
in the optical range have been carried out by Dinshaw et
al. (\cite{Dinshaw}; 
$\theta=9\farcs5$, 
$D>160$ $h_{50}^{-1}$ kpc at $1.7<z<2.1$) and Crotts et
al. (\cite{Crotts}; same results).

At low redshift ($z\la 1.5$), Ly$\alpha$ clouds show a very flat
redshift distribution: only  $16\pm 2$ Ly$\alpha$ lines
with $W_0>0.32$ \AA{} are expected in the wavelength interval covered
by the FOS on the {\it HST} (Weymann et
al.~\cite{Weymann1}). Therefore, the size-estimate uncertainties for
low-redshift Ly$\alpha$ absorbers are large. Studies using FOS spectra of 
QSO pairs have been made by Dinshaw et al. (\cite{Dinshaw2}; 
$\theta=1\farcm29$, $D>600$ $h_{50}^{-1}$ kpc at $0.5<z<0.9$), and more
recently by Petitjean et al. (\cite{Petitjean}; $\theta=36\arcsec$, $D\sim 500$
$h_{50}^{-1}$ kpc at $0.8<z<1.4$). 

All the aforementioned limits on Ly$\alpha$ cloud sizes result from
rather crude models which assume non-evolving and uniform-sized
absorbers. Such simple models might be unrealistic, as
suggested for example by a trend of larger size estimates with larger
LOS separation (Fang et al.~\cite{Fang}) or by the  claim (Fang et
al.~\cite{Fang}, Dinshaw et 
al.~\cite{Dinshaw1}) that the cloud size increased with cosmic
time. However, we note that none of the above claims is confirmed
by  D'Odorico et  al. (\cite{D'Odorico}), using a larger database. 
Clearly, the issues of geometry and size of Ly$\alpha$
absorbers remain open due partly to a lack of size estimates at
low redshift. 
    
In this paper we use a likelihood analysis (\S~\ref{1216_sizes}) that
attempts to constrain the size of Ly$\alpha$ clouds in front of \hs{} A
and B based on simple considerations on shape and evolution. We model
the absorbers either as non-evolving spheres or as filaments. The
technique takes advantage of the information 
provided by line pairs observed in both 
spectra at the same redshift, hereafter 
referred to as ``coincidences'', and Ly$\alpha$ lines observed only in
one spectrum, referred to as ``anti-coincidences''. A likelihood
function is then constructed using an analytic expression for the
probability of getting the observed number of coincidences and
anti-coincidences in each of the two geometries. 

In general, modelling the geometry of Ly$\alpha$ absorbers at low
redshift on the basis of line counting  is
made difficult because the line samples are intrinsically
small, implying a poor statistics. In addition, \hs{} B shows broad absorption 
lines (BALs) caused by several ions with transitions in the UV, so the
effective redshift path for the detection of Ly$\alpha$ lines is even
shorter.

Intervening metal absorption lines of interest are
presented in \S~\ref{1216_sys}. The BAL systems are described in
\S~\ref{sec_BAL}. The conclusions are outlined in \S~\ref{sec_dis}. 

\section{Data reduction}

\subsection{Observations and wavelength calibration}
\label{1216_obs}

UV spectra of HS 1216+5032 A and B were obtained on 1996 November 6
with the FOS on board the {\it HST}. Target acquisition and
spectroscopy were done 
using Grating G270H with the red detector and the $3.\!{}''7\times3.\!{}''7$
aperture. This configuration
yields a spectral resolution of FWHM $=2$ \AA{} and a wavelength coverage from
2222 \AA{} to 3277 \AA{} (Schneider et al. \cite{Schneider}). Total
integration times were 1980 and 10400 s for QSO components A and B,
respectively. The spectrum of image B resulted from the 
variance-weighted addition of four exposures.  
The  signal-to-noise ratios at Ly$\alpha$ emission are S/N $=40$ (A) and $25$  
per $\sim 0.5$ \AA{} pixel, falling down to 15 in the blue part of the A
spectrum, and to near one at the BAL troughs in B. 

\begin{figure*}
      \vspace{0cm}
\hspace{0cm}\epsfig{figure=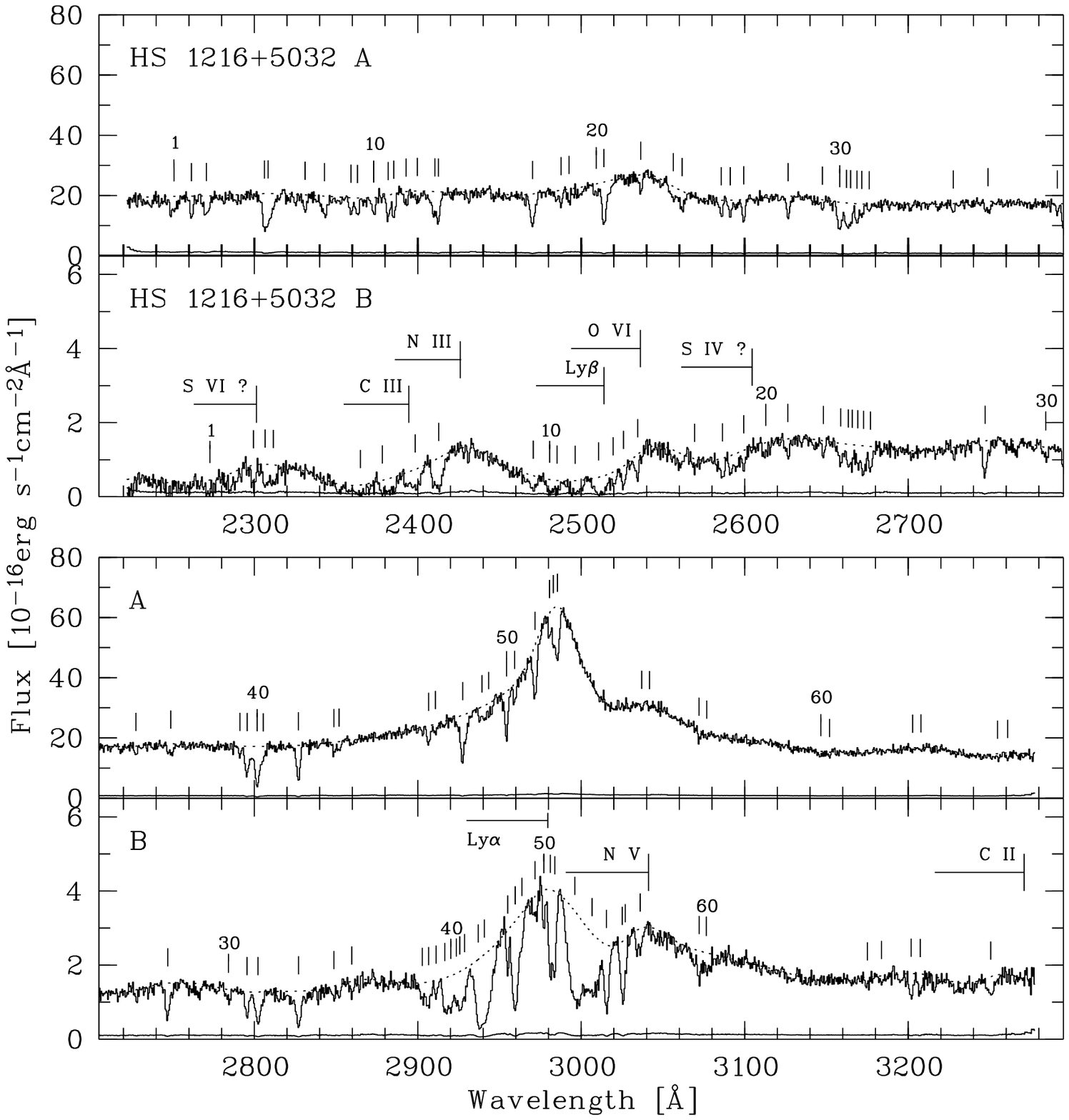,width=17.6cm}
\vspace{0cm}
      \caption[]{
      {\it HST} FOS G270H spectra of HS 1216+5032 A and B and their
      1$\sigma$ errors. The dotted line represents the fitted 
      continuum (\S~\ref{1216_cont}) and the ticks mark the positions
      of   $3  \sigma$ absorption  
      lines (\S~\ref{1216_abs}), with exception of less significant
      \ion{C}{iv} and \ion{Mg}{ii} 
      doublet lines whose identification is supported by the detection
      of the corresponding Ly$\alpha$ line (see \S~\ref{1216_abs} and
      Tables~\ref{tbl-1_1216} 
      and~\ref{tbl-2_1216}). The vertical bars in the B plots indicate
      the expected   position of emission lines at $z_e=1.451$. 
The horizontal bars  correspond to a velocity of $5\,000$
      \kms{} in the QSO rest frame, a region thought to be affected by
      the BALs (see also Fig.~\ref{fig2_1216}). Notice in B the broad
troughs on the blue side of the \ion{S}{iv}, \ion{O}{vi},
\ion{N}{iii}, \ion{C}{iii} and possibly \ion{S}{vi} resonance
      lines.

              }
         \label{fig1_1216}
   \end{figure*}

The flux-calibrated spectra and their associated 1
$\sigma$ errors are shown in Figure~\ref{fig1_1216}.
The dotted lines
represent the continua (\S~\ref{1216_cont}) and the tick-marks indicate
the position of $3 \sigma$ absorption lines (\S~\ref{1216_abs}).

Since we are interested in comparing absorption systems along both LOSs,
we must check for possible small misalignments of the zero-point      
wavelength scale between both spectra. Such differences may arise when
the targets have not been properly centered in the aperture of the
FOS, and/or when -- unlike here -- the science and wavelength
calibration exposures have  
not been performed consecutively. Galactic absorption
lines common to both spectra were then required to appear at the same
wavelength, under the  
assumption that both LOSs cross the same cloud (which should be true
for the small separation between LOSs at $z\approx 0$). There are two
Galactic absorption lines, \ion{Fe}{ii} $\lambda2600$ and
\ion{Mg}{ii} $\lambda2796$, common to both spectra and apparently not
contaminated with intergalactic lines. For these lines 
$\Delta\lambda$(\ion{Fe}{ii} $\lambda2600$)$=-9$ \kms{} and
$\Delta\lambda$(\ion{Mg}{ii} $\lambda2796$)$=+4$ \kms{} are obtained,
which lie within the fitting procedure errors ($\sim10$ \kms) and the
FOS limiting accuracy ($\sim20$ \kms). These small differences imply
that both spectra are well-calibrated; consequently, we made no
corrections to the zero-point wavelength scale.

\subsection{Emission redshifts}
\label{1216_em}

To derive the emission redshift of HS 1216+5032 A  
the Ly$\alpha$ emission peak was fitted with a Gaussian in the region between
2959 and 3005 \AA{}. Excluding
the wavelength  regions with absorption features in the blue wing of the
line yields $z_e({\rm A})=1.4545\pm 0.0001$, where the $1 \sigma$ error comes
from the wavelength uncertainty. For QSO component B, 
a similar analysis is extremely difficult because the Ly$\alpha$
emission line is severely distorted by the Ly$\alpha$ and \ion{N}{v}
BALs. Fitting instead the \ion{N}{v} emission peak between 3016 and
3065 \AA, and using the mean rest-frame wavelength of the doublet
$\lambda_0$(\ion{N}{v})$=1240.15$ \AA, yields $z_e({\rm B})=1.4509\pm
0.0006$. In the following we adopt the nominal values $z_e$(A)$=1.455$ and
$z_e$(B)$=1.451$. 

The redshifts of A and B are not the same within
measurement errors, with QSO component B being blueshifted by $\sim
400$ \kms{}  relative to A. Different emission redshifts are expected in
the physical pair hypothesis, but the measured difference between \hs{} A
and B cannot rule out the possibility that the QSO may still be
gravitationally lensed. This is 
because two different transitions, Ly$\alpha$ and \ion{N}{v}, are being
compared and blueshifts
of a few hundreds \kms{} between high and low-ionization emission lines
are common (e.g., Hamann et al.~\cite{Hamman}; Tytler \&
Fan~\cite{Tytler}). This point is further discussed in
\S~\ref{1216_lensed?}. 

Other prominent emission lines in the spectrum of A are: 
\ion{O}{vi} $\lambda1033$ (blended with Ly$\beta$) and probably
\ion{C}{iii} $\lambda977$ and \ion{S}{iv} $\lambda1062$. 
In the spectrum of B, the BAL troughs allow clear identification of 
only Ly$\alpha$ and \ion{N}{v} $\lambda1240$, but
the ``undulating'' shape of the observed flux for $\lambda<2600$
\AA{} suggests that \ion{S}{vi} $\lambda939$, \ion{C}{iii} $\lambda977$, 
\ion{N}{iii} $\lambda990$, \ion{O}{vi} $\lambda1033$, and
\ion{S}{iv} $\lambda1062$ have broad absorption troughs on the blue
side of the expected QSO rest wavelengths. 

\subsection{Continuum fitting}
\label{1216_cont}

Due to line blending in the Ly$\alpha$ forest, defining a  quasar 
continuum shortward of Ly$\alpha$ emission is not trivial at our
resolution. For B, this difficulty is increased by the BAL profiles. A
power-law shaped continuum $f\propto\nu^{\alpha}$, although an acceptable 
representation in the optical range, is definitively not able to fit even
the UV continuum spectrum of quasar image A, leaving residuals far larger than
3 $\sigma$ at apparently featureless spectral regions.

We decided to fit the UV QSO continuum for each spectrum with cubic
splines, using the following simple and semi-automated fit
algorithm. First, cubic splines are fitted through
a given number of points $(x_i,y_i)$ thought to represent the
continuum, typically 50 in each spectrum. The point ordinates $y_i$ are 
then corrected by an
amount, such that the mean of the differences between the new
ordinate and all the flux values within $1 \sigma$ (defined by the
first fit) in a  6
\AA{} wide spectral window around $x_i$ equals zero. The corrected $(x_i,y_i)$
are used to perform a new fit. This procedure is iterated
until it converges to a reasonable continuum, provided enough
featureless regions are available in the data. The fit routine is robust
in the sense that different start values always lead to the same
continuum, but the spectral windows need to be carefully chosen, 
especially in the B spectrum. The final continuum is shown as dotted
line in Fig.~\ref{fig1_1216}. Note that we have not attempted to model the
QSO intrinsic continuum of B, which is evidently distorted by the
BALs. Division of the flux by this continuum
results in the normalized spectra of HS 1216+5032 A and B.

\section{Absorption line parameters and identifications}
\label{1216_abs}

   \begin{table}

      \caption[]{Absorption Lines in HS 1216+5032 A.}
         \label{tbl-1_1216}
\scriptsize
      \[
         \begin{array}{rccccclrc}
            \hline
            \noalign{\smallskip}
{\rm Line}&\lambda_{\rm obs}&\sigma_{\lambda}&{\rm FWHM}&W_{\lambda}&
\sigma_W&\multicolumn{1}{c}{\rm ID}&\multicolumn{1}{c}{z}&{\rm note}\\
            \noalign{\smallskip}
            \hline
            \noalign{\smallskip}
    1& 2250.88& 0.44& 4.99& 1.27& 0.27& \ion{H}{i}~1215   &   0.85155&\\
    2& 2261.61& 0.12& 2.10& 0.81& 0.15& \ion{H}{i}~972   &   1.32548&\\
    3& 2270.72& 0.19& 4.31& 1.42& 0.20& \ion{H}{i}~1215   &   0.86787&\\
    4& 2306.20& 0.16& 2.22& 0.58& 0.19& \ion{H}{i}~1215   &   0.89706&\\
    5& 2308.44& 0.31& 6.66& 3.29& 0.32& \ion{H}{i}~1215   &   0.89890&\\
    6& 2331.07& 0.16& 2.64& 0.75& 0.14& \ion{H}{i}~1215   &   0.91752&\\
    7& 2342.94& 0.17& 4.21& 1.42& 0.19& \ion{Fe}{ii}~2344  &  -0.00054&1\\
    8& 2359.20& 0.15& 2.45& 0.75& 0.14& \ion{H}{i}~1215   &   0.94066&\\
    9& 2363.08& 0.14& 2.41& 0.79& 0.15& \ion{H}{i}~1215   &   0.94385&10\\
   10& 2373.10& 0.14& 2.46& 0.77& 0.13& \ion{H}{i}~1215   &   0.95210&\\
   11& 2381.84& 0.10& 2.26& 0.99& 0.13& \ion{Fe}{ii}~2382  &  -0.00039&1\\
   12& 2385.30& 0.10& 2.32& 0.99& 0.14& \ion{H}{i}~1025   &   1.32548&6\\
   13& 2392.87& 0.19& 2.61& 0.57& 0.13& \ion{Si}{iv}~1393  &   0.71690&\\
   14& 2399.75& 0.45& 5.76& 0.84& 0.23& \ion{H}{i}~1215   &   0.97402&2\\
   15& 2410.50& 0.28& 6.18& 2.28& 0.23& \ion{H}{i}~1215   &   0.98285&\\
   16& 2412.67& 0.11& 1.58& 0.43& 0.12& \ion{H}{i}~1215   &   0.98464&\\
   17& 2470.24& 0.08& 3.38& 1.82& 0.12& \ion{H}{i}~1215   &   1.03200&3\\
   18& 2487.68& 0.13& 1.69& 0.46& 0.11& \ion{H}{i}~1215   &   1.04635&\\
   19& 2492.54& 0.21& 2.09& 0.46& 0.15& \ion{H}{i}~1025   &   1.43003&\\
   20& 2509.24& 0.28& 3.39& 0.55& 0.15& \ion{H}{i}~1215   &   1.06408&4\\
   21& 2513.90& 0.14& 2.66& 1.60& 0.25& \ion{H}{i}~1215   &   1.06791&\\
   22& 2536.43& 0.12& 2.07& 0.49& 0.09& \ion{H}{i}~1215   &   1.08644&5\\
   23& 2556.38& 0.32& 4.46& 0.86& 0.19& \ion{H}{i}~1215   &   1.10286&\\
   24& 2561.73& 0.18& 3.45& 1.04& 0.15& \ion{H}{i}~1215   &   1.10726&7\\
   25& 2585.71& 0.11& 2.03& 0.64& 0.11& \ion{Fe}{ii}~2586  &  -0.00036&1\\
   26& 2591.15& 0.12& 2.02& 0.65& 0.11& \ion{H}{i}~1215   &   1.13146&\\
   27& 2599.35& 0.07& 1.98& 0.88& 0.10& \ion{Fe}{ii}~2600  &  -0.00032&1\\
   28& 2626.59& 0.08& 2.27& 0.89& 0.10& \ion{H}{i}~1215   &   1.16062&\\
   29& 2647.53& 0.17& 1.74& 0.34& 0.10& \ion{C}{ii}~1334   &   0.98387&\\
   30& 2658.10& 0.12& 3.17& 1.77& 0.18& \ion{C}{iv}~1548   &   0.71690&\\
   31& 2662.31& 1.00& 3.01& 1.20& 0.83& \ion{C}{iv}~1550   &   0.71677&\\
   32& 2664.83& 1.00& 2.88& 1.11& 0.81& \ion{C}{iv}~1548   &   0.72125&\\
   33& 2668.68& 0.11& 2.02& 0.84& 0.15& \ion{C}{iv}~1550   &   0.72087&\\
   34& 2671.66& 0.20& 2.47& 0.62& 0.15& \ion{C}{iv}~1548   &   0.72566&\\
   35& 2676.23& 0.23& 1.92& 0.28& 0.11& \ion{C}{iv}~1550   &   0.72574&\\
   36& 2727.61& 0.16& 1.80& 0.35& 0.10& \ion{H}{i}~1215   &   1.24371&\\
   37& 2748.87& 0.22& 4.18& 0.87& 0.15& \ion{H}{i}~1215   &   1.26119&\\
   38& 2791.18& 0.11& 1.28& 0.30& 0.08& \ion{H}{i}~1215   &   1.29600&\\
   39& 2795.61& 0.07& 3.23& 1.88& 0.13& \ion{Mg}{ii}~2796  &  -0.00027&1\\
   40& 2801.78& 0.09& 3.51& 2.83& 0.18& \ion{Mg}{ii}~2803  &  -0.00063&1\\
   41& 2805.50& 0.27& 2.89& 0.73& 0.17& \ion{H}{i}~1215   &   1.30778&\\
   42& 2826.96& 0.04& 2.43& 1.77& 0.09& \ion{H}{i}~1215   &   1.32543&\\
   43& 2848.74& 0.15& 1.76& 0.41& 0.10& \ion{H}{i}~1215   &   1.34335&\\
   44& 2851.83& 0.24& 2.52& 0.44& 0.13& \ion{H}{i}~1215   &   1.34589&\\
   45& 2906.67& 0.11& 2.27& 0.64& 0.10& \ion{H}{i}~1215   &   1.39100&9\\
   46& 2910.77& 0.26& 2.68& 0.35& 0.11& \ion{H}{i}~1215   &   1.39437&9\\
   47& 2927.54& 0.06& 4.24& 2.33& 0.10& \ion{H}{i}~1215   &   1.40817&\\
   48& 2939.28& 2.20& 4.83& 0.64& 0.52& \ion{C}{iv}~1548  &   0.89852&11\\
   49& 2943.34& 2.05& 4.57& 0.59& 0.53& \ion{C}{iv}~1550  &   0.89799&11\\
   50& 2954.28& 0.05& 2.38& 1.11& 0.07& \ion{H}{i}~1215   &   1.43016&\\
   51& 2959.27& 0.12& 2.08& 0.38& 0.07& \ion{H}{i}~1215   &   1.43427&\\
   52& 2971.79& 0.06& 2.83& 1.04& 0.07& \ion{H}{i}~1215   &   1.44457&\\
   53& 2980.60& 0.18& 1.70& 0.25& 0.07& \ion{H}{i}~1215   &   1.45182&8\\
   54& 2982.93& 0.17& 1.85& 0.34& 0.11& \ion{H}{i}~1215   &   1.45373&8\\
   55& 2985.43& 0.17& 2.76& 0.84& 0.11& \ion{H}{i}~1215   &   1.45579&8\\
   56& 3037.03& 0.38& 1.78& 0.11& 0.07& \ion{C}{iv}~1548   &   0.96166&\\
   57& 3041.81& 0.53& 1.87& 0.09& 0.08& \ion{C}{iv}~1550   &   0.96148&\\
   58& 3072.14& 0.15& 1.46& 0.25& 0.08& \ion{C}{iv}~1548   &   0.98434&\\
   59& 3076.81& 0.23& 1.15& 0.11& 0.07& \ion{C}{iv}~1550   &   0.98405&\\
   60& 3146.54& 0.58& 4.98& 0.49& 0.18& \ion{C}{iv}~1548   &   1.03239&\\
   61& 3151.91& 0.19& 1.13& 0.15& 0.08& \ion{C}{iv}~1550   &   1.03248&\\
   62& 3202.78& 0.56& 2.44& 0.17& 0.12& \ion{C}{iv}~1548   &   1.06872&\\
   63& 3207.82& 0.70& 2.54& 0.15& 0.13& \ion{C}{iv}~1550   &   1.06853&\\
   64& 3254.73& 0.59& 4.53& 0.47& 0.20& \ion{C}{iv}~1548   &   1.10227&\\
   65& 3260.87& 0.24& 1.84& 0.30& 0.12& \ion{C}{iv}~1550   &   1.10275&\\
            \noalign{\smallskip}
            \hline
         \end{array}
      \]
\begin{list}{}{}
\item[1] Absorption in the ISM.
\item[2] Or \ion{Si}{iv} $\lambda1402$ at $z=0.721$.
\item[3] Blended with \ion{H}{i} $\lambda1025$ at $z=1.4082$ if $b>30$ \kms.
\item[4] Or \ion{Si}{ii} $\lambda1260$ at $z=0.984$.
\item[5] Or \ion{Si}{iii} $\lambda1206$ at $z=1.102$.
\item[6] Blended with \ion{H}{i} $\lambda1215$ at $z=0.9616$.
\item[7] Or \ion{Si}{ii} $\lambda1260$ at $z=1.032$.
\item[8] Associated system. Ly$\beta$ is blended with line A21.
\item[9] Lines mimic \ion{C}{iv} doublet but no corresponding
\ion{H}{i} $\lambda1215$ line is detected.
\item[10] Small contribution from \ion{H}{i}
         $\lambda 972$ at $z=1.430$ to $W_\lambda$.
\item[11] Strong blend. The identification is supported by the
          detection of  \ion{H}{i} (A4 and A5) 
\end{list}
         \end{table}



   \begin{table}

      \caption[]{Absorption Lines in HS 1216+5032 B.}
         \label{tbl-2_1216}
\scriptsize
      \[
         \begin{array}{rccccclrc}
            \hline
            \noalign{\smallskip}
{\rm Line}&\lambda_{\rm obs}&\sigma_{\lambda}&{\rm FWHM}&W_{\lambda}&
\sigma_W&\multicolumn{1}{c}{\rm ID}&\multicolumn{1}{c}{z}&{\rm note}\\
            \noalign{\smallskip}
            \hline
            \noalign{\smallskip}
    1& 2272.88& 0.34& 6.60& 6.41& 1.10& \ion{S}{vi}~933     &  1.43511&\\
    2& 2299.56& 0.19& 3.44& 2.57& 0.44& \ion{S}{vi}~944     &  1.43463&\\
    3& 2306.60& 0.46& 3.91& 2.52& 0.73& \ion{H}{i}~1215    &  0.89739&\\
    4& 2311.73& 0.66& 5.61& 3.49& 1.01& \ion{H}{i}~1215    &  0.90161&\\
    5& 2364.96& 0.64& 10.1& 8.49& 1.76& \ion{C}{iii}~977 {\rm BAL}& 1.42058&\\
    6& 2378.32& 0.25& 2.72& 2.62& 0.81& \ion{C}{iii}~977{\rm  BAL}& 1.43426&2\\
    7& 2398.41& 0.32& 6.99& 5.26& 0.74& \ion{N}{iii}~989 {\rm BAL}& 1.42313&\\
    8& 2412.84& 0.24& 5.42& 4.70& 0.69& \ion{N}{iii}~989 {\rm BAL}& 1.35233&3\\
    9& 2470.63& 0.36& 4.05& 2.93& 0.88& \ion{H}{i}~1215    &  1.03232&6\\
   10& 2480.74& 0.47& 3.84& 3.10& 0.82& \ion{H}{i}~1025 {\rm BAL}&  1.41853&\\
   11& 2485.24& 0.55& 4.13& 3.06& 0.94& \ion{H}{i}~1025 {\rm BAL}&  1.42292&\\
   12& 2496.28& 0.35& 8.62& 7.34& 0.93& \ion{O}{vi}~1031 {\rm BAL}&  1.41905&\\
   13& 2510.67& 1.00& 6.48& 6.14& 1.74& \ion{O}{vi}~1031 {\rm BAL}& 1.43299&4\\
   14& 2519.51& 0.39& 4.43& 2.40& 0.63& \ion{O}{vi}~1031 {\rm BAL}&  1.44156&\\
   15& 2525.86& 0.12& 2.62& 1.98& 0.27& \ion{O}{vi}~1037 {\rm BAL}&  1.43429&\\
   16& 2534.59& 0.25& 1.70& 1.12& 0.33& \ion{O}{vi}~1037 {\rm BAL}&  1.44270&\\
   17& 2569.40& 0.28& 2.84& 0.84& 0.26& \ion{S}{iv}~1062 {\rm BAL?}& 1.41789&\\
   18& 2586.38& 0.21& 3.28& 1.78& 0.32& \ion{Fe}{ii}~2586   & -0.00010&1,5\\
   19& 2599.43& 0.13& 2.42& 1.10& 0.22& \ion{Fe}{ii}~2600   & -0.00029&1\\
   20& 2612.86& 0.33& 4.21& 1.01& 0.25& \ion{H}{i}~1215    &  1.14932&\\
   21& 2626.39& 0.21& 3.18& 1.01& 0.20& \ion{H}{i}~1215    &  1.16045&\\
   22& 2648.20& 0.17& 1.81& 0.57& 0.16& \ion{C}{ii}~1334    &  0.98437&\\
   23& 2658.72& 0.21& 3.10& 1.24& 0.24& \ion{C}{iv}~1548    &  0.71730&\\
   24& 2663.43& 1.09& 2.97& 1.21& 0.84& \ion{C}{iv}~1550    &  0.71749&\\
   25& 2665.83& 0.79& 2.34& 0.86& 0.82& \ion{C}{iv}~1548    &  0.72190&\\
   26& 2669.14& 0.26& 2.40& 0.96& 0.35& \ion{C}{iv}~1550    &  0.72117&\\
   27& 2672.69& 0.20& 3.43& 2.04& 0.43& \ion{C}{iv}~1548    &  0.72633&\\
   28& 2677.01& 0.15& 2.50& 1.31& 0.22& \ion{C}{iv}~1550    &  0.72625&\\
   29& 2747.05& 0.10& 3.49& 2.21& 0.18& \ion{H}{i}~1215    &  1.25970&11\\
   30& 2784.24& 0.29& 2.97& 0.72& 0.22& \ion{Si}{iv}~1402   &  0.98482&\\
   31& 2795.57& 0.10& 2.42& 1.39& 0.18& \ion{Mg}{ii}~2796   & -0.00028&1\\
   32& 2802.26& 0.55& 3.68& 2.57& 0.69& \ion{Mg}{ii}~2803   & -0.00045&1\\
   33& 2827.11& 0.10& 3.39& 2.69& 0.27& \ion{H}{i}~1215    &  1.32556&\\
   34& 2848.69& 0.30& 2.88& 0.70& 0.23& \ion{H}{i}~1215    &  1.34331&\\
   35& 2859.71& 0.23& 2.49& 0.62& 0.17& \ion{H}{i}~1215    &  1.35237&\\
   36& 2902.79& 0.47& 3.39& 1.04& 0.33& \ion{H}{i}~1215    &  1.38781&\\
   37& 2906.60& 0.30& 3.20& 1.45& 0.32& \ion{H}{i}~1215    &  1.39094&7\\
   38& 2911.22& 0.16& 1.91& 0.59& 0.15& \ion{H}{i}~1215    &  1.39475&7\\
   39& 2916.57& 0.33& 2.94& 1.49& 0.54& \ion{Mg}{ii}~2796   &  0.04299&8\\
   40& 2920.25& 0.47& 4.67& 2.64& 0.92& \ion{Mg}{ii}~2796   &  0.04431&8\\
   41& 2923.58& 0.19& 1.22& 0.31& 0.23& \ion{Mg}{ii}~2803   &  0.04282&8\\
   42& 2925.69& 0.48& 2.94& 1.64& 0.94& \ion{Mg}{ii}~2803   &  0.04357&8\\
   43& 2928.69& 1.13& 3.53& 1.42& 0.77& \ion{H}{i}~1215 ?  & 1.40912&\\
   44& 2936.95& 0.14& 3.29& 1.06& 0.24& \ion{H}{i}~1215 {\rm BAL}&  1.41591&\\
   45& 2940.68& 0.24& 9.41& 8.06& 0.37& \ion{H}{i}~1215 {\rm BAL}&  1.41898&\\
   46& 2954.99& 0.07& 1.33& 0.56& 0.09& \ion{H}{i}~1215    &  1.43075&\\
   47& 2959.60& 0.06& 4.41& 3.61& 0.18& \ion{H}{i}~1215 {\rm BAL}&  1.43454&\\
   48& 2963.76& 0.18& 2.18& 0.48& 0.12& \ion{H}{i}~1215 {\rm BAL}&  1.43796&\\
   49& 2971.75& 0.26& 3.33& 0.52& 0.13& \ion{H}{i}~1215    &  1.44454&\\
   50& 2977.11& 0.09& 1.50& 0.45& 0.08& \ion{H}{i}~1215    &  1.44895&\\
   51& 2981.12& 0.12& 2.66& 1.66& 0.17& \ion{H}{i}~1215    &  1.45224&9\\
   52& 2983.79& 0.11& 2.23& 1.28& 0.14& \ion{H}{i}~1215    &  1.45444&9\\
   53& 2996.12& 0.29& 9.16& 6.52& 0.46& \ion{N}{v}~1238 {\rm BAL}&  1.41853&\\
   54& 3006.67& 0.41& 11.4& 6.81& 0.69& \ion{N}{v}~1242 {\rm BAL}&  1.41926&\\
   55& 3015.40& 0.06& 3.94& 2.86& 0.14& \ion{N}{v}~1238    &  1.43409&\\
   56& 3025.12& 1.00& 2.58& 1.52& 0.67& \ion{N}{v}~1242    &  1.43411&\\
   57& 3026.94& 1.86& 2.43& 0.66& 1.25& \ion{N}{v}~1238    &  1.44340&\\
   58& 3036.12& 0.46& 1.47& 0.34& 0.31& \ion{N}{v}~1242    &  1.44296&\\
   59& 3072.15& 0.15& 1.76& 0.53& 0.17& \ion{C}{iv}~1548    &  0.98434&\\
   60& 3076.58& 0.54& 6.86& 1.50& 0.43& \ion{C}{iv}~1550    &  0.98390&\\
   61& 3175.03& 0.40& 2.98& 0.45& 0.19& \ion{Mg}{ii}~2796   &  0.13542&\\
   62& 3183.64& 0.55& 2.91& 0.35& 0.23& \ion{Mg}{ii}~2803   &  0.13558&\\
   63& 3201.81& 0.15& 3.13& 1.21& 0.18& \ion{C}{iv}~1548    &  1.06809&\\
   64& 3207.46& 0.20& 3.36& 1.06& 0.20& \ion{C}{iv}~1550    &  1.06830&\\
   65& 3250.52& 0.58& 5.02& 1.53& 0.41& \ion{C}{ii}~1334 {\rm BAL}&1.43570&10\\
            \noalign{\smallskip}
            \hline
         \end{array}
      \]
\begin{list}{}{}
\item[1] Absorption in the ISM.
\item[2] Probably blended with Ly-$\alpha$. 
\item[3] Blended with Ly$\alpha$ at $z=0.984$.
\item[4] Blended with  \ion{H}{i} $\lambda1025$ BAL.
\item[5] Also \ion{S}{iv} $\lambda1062$ BAL because the line
is too strong compared with IS \ion{Fe}{ii} $\lambda2600$.
\item[6] Or \ion{H}{i} $\lambda1025$ at $z=1.4082$ if $b>30$ \kms.
\item[7] Lines mimic \ion{C}{iv} doublet but no corresponding
\ion{H}{i} $\lambda1215$ line is detected.
\item[8] The fit procedure is not able to resolve the second doublet
lines; therefore, the doublet ratios are probably not real.
\item[9] Associated system. Ly$\beta$ falls in the \ion{O}{vi} BAL troughs.
\item[10] Blended with \ion{C}{ii}$^{*}~\lambda 1035$. See
\S~\ref{sec_BAL} for the classification as BAL system. 
\item[11] No line at $3\sigma$ can be identified with Ly$\beta$.
\end{list}
         \end{table}

Tables~\ref{tbl-1_1216} and~\ref{tbl-2_1216} list absorption lines found at the
$3 \sigma$ level in HS 1216+5032 A and B, respectively. Lines in the
normalized spectra were fitted 
with Gaussian profiles without constraints in central
wavelength, width  or amplitude. In the case of obvious blends two or more
Gaussian  components were fitted simultaneously.

The fit routine
attempts to minimize $\chi^{2}$ between model and data, considering $1
\sigma$ flux errors (flat-fielding, background
subtraction and
photon statistics noise); but the uncertainties introduced by the 
continuum placement are not included. 

An inherent feature of
$\chi^{2}$ minimization is the non-uniqueness of the solution due to
the eventual presence of more than one minimum in
the parameter space. In those cases, the fit  can lead not only to
wrong parameter 
estimations but also to underestimated parameter errors. To handle
with this problem, instead of
defining a rejection parameter we simply performed several fit trials
taking different wavelength ranges to see how robust a particular
solution was. In general, there were no significant changes in the
resulting parameters due to a different choice of the fitting
interval, thus giving us confidence on the results. However, some fits yield
$\chi^{2}_\nu<1$. Assuming that the  $1 \sigma$ flux errors are not
overestimated, such behaviour can occur by chance, especially when
too few pixels are considered in the fit.

The equivalent-width error estimates are generally larger
than the propagated error if the flux values are integrated along a
line. This is because the former errors consider the correlation
between all three free parameters. 
This is not the case for multicomponent fits, where the errors in
equivalent width may be overestimated  since the fit 
algorithm does not consider that the strengths and widths of neighbour
lines in blends are anti-correlated. 

Concerning the line widths, note that several FWHM values in
Tables~\ref{tbl-1_1216} and~\ref{tbl-2_1216} are slightly larger than the
nominal width of the line spread function (LSF), $2$ \AA. This is caused by
the difficulty in resolving absorption structures separated by 
less than $\sim200$ \kms, since single-line fits of closely lying line
complexes (blends) will inevitably lead to large widths. On the other
hand, in few cases the   
measured lines are narrower than the LSF. While this might partly be
caused by a too low placement of the continuum near emission lines and
BAL troughs in the B spectrum, it is certainly not  the case for all the
narrow lines in the A spectrum. In the latter case fitting {\it
weak} lines also leads to low FWHM values (with large associated
errors), thus reflecting the limitations of the $\chi^2$ minimization
using too few pixels.

BAL profiles in the B spectrum have been fitted with multicomponent
Gaussians to better  constrain their equivalent
widths, but these fits do not necessarily represent the actual nature
of the BAL profiles (see \S~\ref{sec_BAL}). 
For consistency, Tables~\ref{tbl-1_1216} and~\ref{tbl-2_1216} include 
the Gaussian parameters of BAL profiles.

Absorption lines were identified manually, using vacuum wavelengths and
oscillator strengths taken from 
Verner et al. (\cite{Verner}). The identification of lines is 
not easy because of  the low resolution and the lack of optical data; 
but it is especially difficult  
in the B spectrum because narrow intergalactic
lines are hidden among the BALs. We looked first for
interstellar absorption lines in both spectra. The \ion{Mg}{ii}
$\lambda2796,2803$ doublet and the \ion{Fe}{ii} lines at 2586 \AA{}
and 2600 \AA{} are detected in both spectra; the \ion{Fe}{ii} lines at
2344 \AA{} and 2382 \AA{} are
detected at $3 \sigma$ only in the A spectrum (while in the B spectrum
\ion{Fe}{ii} $\lambda2382$ would appear at the position of BAL
\ion{C}{iii} $\lambda977$). The next step was to search for
\ion{C}{iv} systems and  
their associated Ly$\alpha$ and prominent metal
lines. In addition, two \ion{Mg}{ii} systems at $z=0.04$
(\S~\ref{1216_mgii}) and $1.14$
could be identified in the spectrum of B through the positions of the
doublet lines and their relative strengths. Due to their secure
identification, \ion{C}{iv} $\lambda1548,1550$ and
\ion{Mg}{ii} $\lambda2796,2803$ lines with ${\rm SNR}<3$ have been
retained in  
Tables~\ref{tbl-1_1216} and~\ref{tbl-2_1216}, which otherwise contain only 
lines with ${\rm SNR}>3$. Ly$\alpha$ lines with redshifts of
$1.16<z<z_e$ were identified 
through the corresponding Ly$\beta$ and Ly$\gamma$ lines, if
present. Lines A45, A46, 
B37, and B38  mimic a \ion{C}{iv} doublet at
$z=0.88$, but since no significant absorption by \ion{H}{i} is detected at
this redshift, they have been counted as Ly$\alpha$ lines. 
The remaining lines have been
considered to be Ly$\alpha$ lines and will be discussed in
\S~\ref{1216_sizes}.


\section{Metal absorption systems in \hs{} A and B}
\label{1216_sys}

We now give a description of the most remarkable metal absorption
systems observed in \hs{} A and B: a \ion{Mg}{ii} system at $z=0.04$,
and a strong \ion{C}{iv} system at $z=0.72$.

\subsection{The \ion{C}{iv} systems at $\mathit{z=0.72}$ in HS 1216+5032 A and B}
\label{1216_civ}

Three strong \ion{C}{iv} systems at $z=0.717$, $0.721$ and $0.726$  are
identified through the
$\lambda\lambda1548,1550$ doublets at $\lambda\sim 2660$ \AA{} in 
both spectra of HS 1216+5032. Although some of the \ion{C}{iv} lines
are under the $3 \sigma$ significance level, the well-determined
redshifts of the doublet components make their
identification unambiguous; consequently, all six  \ion{C}{iv} lines
appear in Tables~\ref{tbl-1_1216} and~\ref{tbl-2_1216}. 

The rest frame
equivalent widths of the $\lambda1548$ lines range between $0.36$
and $1.03$ \AA{} in the A spectrum, and between 0.72 and 1.19 \AA{} in
B. The large column densities implied by these line strengths, 
make this system a firm candidate for a Lyman-limit system
(Sargent, Steidel \&{} Boksenberg~\cite{Sargent}). 
Also, the large line widths in A and B, FWHM $>2$ \AA, suggest that these
\ion{C}{iv} systems will reveal several
narrower components at higher resolution. 

No further metal lines are found at these redshifts. Line A13 could be
identified with \ion{Si}{iv} $\lambda1393$ at $z=0.717$, but two
strong Ly$\alpha$ lines at $\lambda=2410$ \AA{} make the detection of
the second doublet line impossible. No
transitions by low-ionization species (e. g., \ion{C}{ii} $\lambda1334$)
are detected. In the low-resolution optical spectrum of HS
1216+5032 B (see Fig. [3] in Hagen et al.~\cite{Hagen}), a significant
absorption feature at $\lambda=4816$ \AA{} could 
tentatively be identified with a \ion{Mg}{ii} $\lambda2796,2803$
doublet associated to these \ion{C}{iv} systems, but no absorption
feature is seen at this wavelength in the optical spectrum of A. 

The small velocity differences of 
$\Delta v_{\rm A-B}\approx -68\pm28$ \kms, $-52\pm31$ \kms, and
$-116\pm30$ \kms{} at 
$z=0.717$, $0.721$ and $0.726$, respectively, between the \ion{C}{iv}
lines in A and the corresponding ones in B suggest strongly that
these \ion{C}{iv} systems are physically associated. Therefore, the
large velocity span of roughly $1500$ \kms{} along each LOS can only
be explained if the gas is virialized by 
a cluster of galaxies. The virial mass within the radius given by the half
separation between the LOSs is derived to be $\sim 10^{13}$
M$_\odot$. Consequently, if \ion{C}{iv} systems arise in the extended
halos of galaxies (Bergeron \&{} Boisse~\cite{Bergeron}), the present
data gives evidence that the \ion{C}{iv} absorbers at $z=0.72$ in HS
1216+5032 A 
and B arise either in the highly ionized intra-group gas of a galaxy
cluster, or in small structures associated with the cluster.

The whole absorption complex shows an asymmetry in the
sense  that the weakest system in spectrum A has the same redshift as
the strongest in B: the ratio of the $\lambda1548$ line's equivalent width
in A to that in B varies
between 1.4 and 3.3. If \ion{C}{iv} systems in A and B with similar
redshifts are physically associated, then the different line strengths have
direct implications for the size of these absorbers. This is  because
such gas inhomogeneities imply that the LOSs sample the clouds on spatial
scales similar to the transverse cloud sizes. 
The interpretation remains valid if the lines in the present FOS spectra 
result from many unresolved velocity components, 
because in that case stronger lines result from more 
narrow components than weaker 
lines do, a picture that is still compatible with gas
inhomogeneities. All these three 
\ion{C}{iv} absorption systems should therefore arise in clouds with {\em 
characteristic} transverse lengths of $\sim 75~h_{50}^{-1}$kpc, larger than
the statistical lower limits of $\sim 30~h_{50}^{-1}$kpc derived by Smette
et al. (\cite{Smette}) for \ion{C}{iv}  
absorbers at $z\ga1.5$.

Alternatively, it is still
possible that the \ion{C}{iv} absorption is correlated in redshift, but
occurs in distinct, separated structures. At higher resolution, Rauch
(\cite{Rauch1}) has shown that in gas associated with metal
systems density gradients on sub-kpc scales are
not uncommon. The logical conclusion would be  that C\,{\sc iv}
absorbers are composed of a large number of small cloudlets. However,
if that is the case of the $z=0.72$ absorbers, the correlation in
redshift between lines in A and B is difficult to explain without
invoking cloudlets aligned along a filamentary or sheet-like
structure. 

\subsection{A \ion{Mg}{ii} System at  $\mathit{z=0.04}$   in HS
1216+5032 B?}
\label{1216_mgii}

There is a strong and blended feature at $\lambda\sim2920$ \AA{}
in the B spectrum for which we do not find any plausible
identification with other observed metal systems. One possible
identification is absorption by  two \ion{Mg}{ii}
$\lambda\lambda2796,2803$ doublets at $z=0.043$
and $z=0.044$. Although the Gaussian profile fit is not able to resolve the
single lines in the red trough of the absorption feature, the
identification is supported by a good match between the line positions
and profiles of the doublet. The red wing of the whole 
absorption feature could be blended with a Ly$\alpha$ line because it
coincides quite well with a strong 
Ly$\alpha$ line at $z=1.4$ in A (A47). 
However, no \ion{Fe}{ii} lines are
detected at this redshift, nor further 
transitions by low-ionization species. The  \ion{Mg}{i} $\lambda 2852$
line is possibly present, but the match in wavelength with line B50,
$\Delta \lambda \approx 1.4$ \AA, is not very good.  
Under the \ion{Mg}{ii}-hypothesis, the total rest-frame equivalent
width of the stronger 
\ion{Mg}{ii} doublet component would be  $W_0=3.97$ \AA{}, which is somewhat
larger than 
typical values found in gas associated with damped Ly$\alpha$ (DLA) systems
at high redshift (Lu 
et al.~\cite{Lu}) or in the Milky Way (Savage et
al.~\cite{Savage}). This could be explained if the lines are indeed made
up of several narrower 
components, as usually seen in DLA systems. The redshift difference
between both \ion{Mg}{ii} systems implies a
velocity span of roughly 300 \kms, also typical of DLA 
systems. Therefore, there is some evidence that these \ion{Mg}{ii}
lines might be associated with a DLA system at $z=0.04$.


Regardless of the DLA-system interpretation, however, even the absence
of \ion{Fe}{ii} and -- possibly -- also 
\ion{Mg}{i} lines associated with \ion{Mg}{ii} of such a strength is
difficult to explain considering the incidence of the former ions in
\ion{Mg}{ii}-selected samples (e.g., Bergeron
\& Stasinska~\cite{Bergeron2}; 
Steidel \& Sargent~\cite{Steidel}). Consequently, 
an alternative identification of the $\lambda\sim2920$ \AA{}
feature as \ion{H}{i} $\lambda 1215$ BAL at $v\sim -6\,000$ \kms{} 
(see \S\ref{sec_BAL} and Figure~\ref{fig2_1216}) must be considered as
well. The absence of a corresponding \ion{H}{i} $\lambda 1025$ BAL
profile, however, is also in this 
case remarkable, but could be explained if the absorber does not
completely cover the continuum source (see below).

\section{The BAL systems in HS 1216+5032 B: Some Qualitative Inferences}
\label{sec_BAL}

\begin{figure}[t]
\centering
      \vspace{0cm}
\hspace{0cm}\epsfig{figure=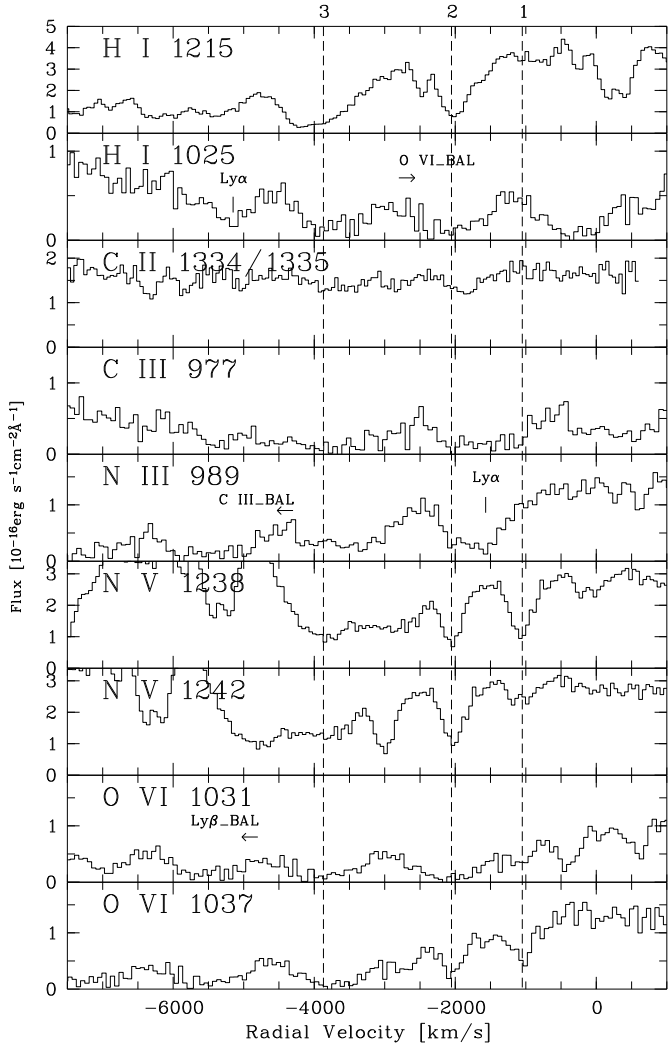,width=8.8cm}
\vspace{0cm}
      \caption[\hs{} B: BALs in velocity space]{
Broad absorption lines in HS 1216+5032 B plotted in velocity relative
to $z_e=1.451$.
              }
         \label{fig2_1216}
   \end{figure}

Three BAL systems are observed in the UV spectrum of HS
1216+5032 B 
(see Fig.~\ref{fig2_1216}, where the 
system redshifts have been arbitrarily numbered 1, 2, and 3). 
Absorption by \ion{H}{i}, \ion{C}{iii}, \ion{N}{iii}, 
and possibly \ion{S}{iv} is observed in at least two of the systems, 
while \ion{O}{vi} and \ion{N}{v} are present in all three systems. 
\ion{C}{ii} is observed only in system 2,
and \ion{N}{iii} in systems 2 and 3. 

Many features distinguish these 
systems  from the most commonly observed  BAL systems (e.g.,
Turnshek et al.~\cite{Turnshek}): (1) the lines  
are particularly weak and several line profiles are
not distorted by/or blended with other BALs; (2) the maximum outflow
velocity $v \sim 5\,000$ \kms{} is small compared with typical
BALQSOs, which exhibit terminal velocities of several $10^4$ \kms
(Turnshek~\cite{Turnshek1}); (3)
\ion{C}{ii} is present (see  
Wampler et al.~\cite{Wampler} for other BAL QSO spectrum with 
singly-ionized species; Arav et al.~\cite{Arav}); (4) the
strength of \ion{H}{i} in systems 2 and 3 
decreases more slowly at the {\it red} edge of the troughs 
(Turnshek~\cite{Turnshek1}). 

Also remarkable is the undulating shape of the absorbed
continuum for $\lambda<2700$ \AA{} (see Fig.~\ref{fig1_1216}). The
position  of
the flux depressions coincides quite well with the
blue wing of expected emission lines by \ion{C}{iii}, \ion{N}{iii},
Ly$\beta$, and \ion{O}{vi}. This
coincidence seems to suggest 
that the absorption is indeed dominated by very broad line components,
which determine the continuum shape, superimposed to the narrower 
components, here labeled as systems 1, 2 and 3. 
The same effect, although less remarkable, is observed for \ion{S}{iv}
and \ion{S}{vi}, thus giving evidence for absorption by these ions  
associated with the BAL phenomenon (also reported for another QSO by  Arav et
al.~\cite{Arav1}).

It is customary to define BALs as a continuous absorption with outflow
velocities larger than $\sim 3\,000$ \kms{} from the emission redshift
(Weymann et al.~\cite{Weymann2}) to make a distinction between BAL
systems and ``associated systems''. However, as pointed out by Arav et
al. (\cite{Arav}), such definition does not hold any physical meaning.
In our case, systems 1 and 2 should be then classified as associated
systems but, due to our poor resolution, it is not possible to
establish whether the line profiles  are produced by continuous
absorption or whether they 
are made of several narrower velocity components (as observed in
associated systems). The metal lines in these systems (e.g. the \ion{N}{v}
and \ion{O}{vi} doublet lines) are relatively narrow and their widths
could be dominated by the instrumental profile (for \ion{H}{i} in
system 2, however, the situation is less clear). Therefore, the
classification of systems 1 and 2 as BAL systems must be considered  solely
instrumental.  

High-resolution spectra of BAL QSOs (Hamann et al.~\cite{Hamman})
show that BAL profiles do not necessarily result from an ensemble
of discrete narrow, unresolved lines, but as part of a mixture
of a continuously accelerated outflow and overlapping narrow
components with different non-thermal velocity dispersions. For the
systems in \hs, it can be assumed that a similar mixture of line widths is
present, with the broad components dominating over the narrow
ones. In that case, these line profiles should not look so different 
at higher resolution (with exception, maybe, of systems 1 and 2).

Ionization models have shown that BAL clouds span a range of densities
and/or distances from the ionizing source (Turnshek et
al.~\cite{Turnshek}; Hamann et
al.~\cite{Hamman1}). Fig.~\ref{fig2_1216}
shows that, while the high-ionization species 
\ion{N}{v} and \ion{O}{vi} are present in all three systems in \hs,
\ion{H}{i}, \ion{C}{ii} and doubly-ionized species are absent in system
1. A possible explanation is  that the ionization conditions could change
with outflow velocity, with higher ionization level for redshifts 
closer to $z_e$. However, a quantitative study of the ionization
conditions in these BAL clouds with the present data is made difficult
by our inability to reliably establish the
continuum level at the BAL troughs and thus
determine column densities (besides the fact that nonblack
saturation of the line profiles might be present; see Arav et
al.~\cite{Arav1}). 

Nevertheless, a more 
quantitative study of the BAL 
phenomenon in \hs{} B should be possible using  higher resolution {\it
HST} spectra  to better estimate 
the QSO continuum and to identify spurious lines at the BAL
troughs. In particular, BAL system 2 can be studied in more detail 
because it shows more ions and less contamination by narrow
absorption lines at the lines lying in the Ly$\alpha$ forest. On the
other hand, alone medium-resolution optical spectra should
considerably improve our knowledge of the BAL systems in \hs{} through the line
profiles of \ion{C}{iv}, \ion{C}{iii}, and possibly \ion{Mg}{ii},
which we suspect to be present given the presence of \ion{C}{ii}. 
The width of the trough in system 2 is sufficiently small
to study nonblack saturation effects via doublet ratios provided the
emission line profiles  can be determined (e.g., at
the position of the \ion{C}{iv} BAL). Thus, corrected column
densities for system 2 appear as very suitable to reliably examine
photoionization models because 
the radiation fields can be constrained by the
presence of low-ionization species.

\section{Ly$\alpha$ absorption systems}
\label{1216_sizes}

We now describe the Ly$\alpha$ forest lines observed in the spectra of
\hs~A and 
B. The projected separation between LOSs samples well the expected
cloud sizes of hundreds of kpc if the QSO pair is real (see next
subsection). However, the effective redshift path 
where Ly$\alpha$ lines can in principle be detected is blocked out by the broad
absorption lines in the B spectrum, so the line samples will be small. 

\subsection{Lensed or Physical Pair?}
\label{1216_lensed?}

The nature of the double images in \hs{} is not clearly established
yet. 
Maybe the strongest arguments
favoring \hs{} AB as a physical-pair instead of a gravitational lens
origin are: (1) BALs are observed only in the spectrum of the B 
image;\footnote{
Curiously, another QSO pair separated by $9\arcsec$, Q1343+266, also
shows BALs in only one spectrum. On the basis of the irregular 
flux ratio in the optical range, Crotts et al. (\cite{Crotts}) find
evidence that this is a genuine QSO pair. From the 
optical spectra of HS 1216+5032 AB (Hagen et al.~\cite{Hagen}),
however, a similar argument is less clear. 
}
 (2) no extended, luminous object is detected between (or even
close to) the QSO images down to $m_R=22.5$; and (3) gravitationally
lensed QSO pairs normally have image separations $\theta\la 3\arcsec$
(e.g., Kochanek, Falco \& Mu\~noz~\cite{Kochanek}).  
In addition, the absence of a trend for equivalent widths of {\em intervening}
Ly$\alpha$  lines common
to both spectra to become similar at redshifts closer
to $z_e$ (see Fig.~\ref{fig3_1216} and Table~\ref{tbl-4_1216}) also
argues for a physical-pair; however, this argument is 
weak as the sample of common lines is too small. Unfortunately, 
for reasons given in \S~\ref{1216_em}, it is not possible to establish
whether there are differences between A and B in the emission line
redshifts and line shapes in the spectral region covered  by the FOS.  

The intriguing point here is that two $z\approx z_e$ Ly$\alpha$
systems are
detected in both spectra at almost identical redshifts to within 
$\sim 90$ \kms{} (see \S~\ref{1216_assoc} for more details). Under the
physical-pair hypothesis, these absorbers should
extend over transverse scales greater than $80~h_{50}^{-1}$kpc. If, on the
other hand, 
\hs{} AB is a gravitationally lensed pair, the detection of these
systems in both spectra can easily be explained by the fact that the
separation between light paths approaches zero. 

In consequence, the gravitational lens nature of \hs{} cannot be yet ruled
out. Let us note that
in a simple lens geometry the virial mass implied by the velocity span
of the \ion{C}{iv} absorbers at $z=0.72$ (see \S~\ref{1216_civ}) is
consistent with the 
deflector mass required to produce an angular separation between QSO
images of $\theta=9\arcsec$. In such a model the separation between light
paths at the source position is derived to be $\theta_s=17$\arcsec. 
Therefore, confirmation of  HS 1216+5032 as
a gravitationally lensed QSO would have dramatic consequences for our
understanding of the BAL phenomenon. As pointed out
by Hagen et al. (\cite{Hagen}), if HS 1216+5032 AB were indeed the
mirror images of one unique QSO, the coverage fraction of BAL clouds 
would have to be very small. Incomplete coverage
of the continuum source or the BLR in BAL
clouds  have been confirmed using high-resolution spectra 
(Hamann et al.~\cite{Hamman}; Barlow \& Sargent~\cite{Barlow}). In fact,
the cloud sizes derived from BAL variability  could be $10^4$ times
smaller than the distances to the source derived from photoionization
models (Hamann et al.~\cite{Hamman1}), implying subtended angles  not much
larger than $\theta_s$. Therefore, the scenario in which only one light
path crosses the BAL clouds is not unrealistic, and the non-BAL
spectrum of HS 1216+5032 A can also be considered if the QSO pair is
lensed.\footnote{Let us mention, however, that in the lensed-pair
hypothesis the fact that BAL
absorption is produced by {\it many} clouds (Hamann et
al.~\cite{Hamman}) lessens considerably the
probability that LOS A does not encounter neither such clouds.} 

In summary, there are reasons to believe that \hs{} is a binary QSO,
but a gravitational lens origin of the double images cannot yet be 
excluded. A definitive assessment of this issue must await medium
resolution optical spectra to better derive emission redshifts and to
establish differences in the continuum or emission
lines. Alternatively, deep infrared or radio observations would help
determining the nature of the double QSO images. The non-detection of
the lensing galaxy/cluster would be a strong proof for the binary
QSO hypothesis, since wide separation ($\theta>3\arcsec$) ``dark
lenses'' enter in conflict 
with current models of structure formation (Kochanek, Falco \&
Mu\~noz~\cite{Kochanek}), which predict few wide separation
gravitational lenses (e.g., Kochanek~\cite{Kochanek1}).

Throughout this
section we will assume that \hs~AB is a physical pair.  
The proper separation $S(z)$ between LOSs obeys then 
the relation $S=\theta~D_{\rm oc}$, where $\theta$ is the angular
separation between the A and B QSO images, and  $D_{\rm oc}$ is the
angular-diameter distance between the observer and a cloud at redshift
$z$. For the redshift range of interest $S\simeq80$ \hkpc. 

\subsection{Line List}
\label{1216_lya}

   \begin{figure}[t]
      \vspace{0cm}
\hspace{0cm}\epsfig{figure=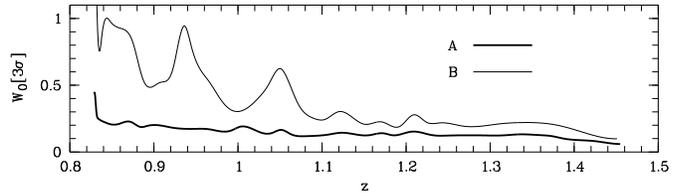,width=8.8cm}
\vspace{0cm}
      \caption[Detection limits for Ly$\alpha$ lines 
in \hs]{ 
Three $\sigma$ rest-frame detection limits of Ly$\alpha$ lines 
in \hs A (thick line) and B.
              }
         \label{fig23}
   \end{figure}

Table~\ref{tbl-4_1216} lists Ly$\alpha$ lines detected at the $3 \sigma$
level in HS 1216+5032 A and B. Figure~\ref{fig23} shows $3 \sigma$
detection thresholds as a function of redshift. 
Out of 36 lines in the spectrum of A, 
a total of 23 lines within $z=0.85$ to $1.43$ have $W_{\rm 0}({\rm
A})\ge0.24$ \AA. This redshift interval excludes lines within $3\,000$
\kms{} from the QSO redshift to avoid the ``proximity effect'' 
(Bechtold~\cite{Bechtold}). The derived number density is 
$\log(dN/dz)=1.58^{+0.09}_{-0.10}$ at $\log(1+z)=0.33\pm0.06$, in very
good agreement with the results of the {\it HST} Key Project (Weymann
et al. \cite{Weymann1}). The scarce
sample of lines in B is clearly explained by the shorter redshift path
allowed by the BAL profiles.

   \begin{table*}[t]
      \caption[Ly$\alpha$ Lines in HS 1216+5032]{Ly$\alpha$ Lines with
      $W>3\sigma_W$ in HS 1216+5032 A and B.} 
         \label{tbl-4_1216}
\scriptsize
      \[
         \begin{array}{crcccrcccrr}
            \hline
            \noalign{\smallskip}
\multicolumn{4}{c}{\rm A} &\multicolumn{4}{c}{\rm
      B}&&\multicolumn{2}{c}{\rm A-B}\\\noalign{\smallskip}
{\rm Line}&\multicolumn{1}{c}{W_0 ({\rm \AA})}&{\rm SNR}&z&{\rm
      Line}&\multicolumn{1}{c}{W_0 ({\rm \AA})}&{\rm
         SNR}&z&{\rm Remarks}&\multicolumn{1}{c}{\Delta v ({\rm
      km~s}^{-1})}&\multicolumn{1}{c}{\Delta W_0~({\rm \AA})}\\ 
            \noalign{\smallskip}
            \hline
            \noalign{\smallskip}
%

  1& 0.68& 4.8  & 0.85155&    &       &    &         & 1  &           & \\
  3& 0.76& 7.0  & 0.86787&    &       &    &         & 2  &           & \\
  4& 0.31& 3.1  & 0.89706&   3&  1.33 & 3.5&  0.89739& 3  & -52\pm63  &-1.02\pm0.40\\
  5& 1.73& 10.2 & 0.89890&    & < 0.47&    &         & 3  &	      &		   \\
   & < 0.19&    &        &   4&  1.84 & 3.5&  0.90161& 4  &	      &		   \\
  6& 0.39& 5.3  & 0.91752&    &       &    &         & 1  &	      &		   \\
  8& 0.39& 5.2  & 0.94066&    &       &    &         & 1  &	      &		   \\
  9& 0.41& 5.5  & 0.94385&    &       &    &         & 2  &	      &		   \\
 10& 0.39& 5.7  & 0.95210&    &       &    &         & 1  &	      &		   \\
 14& 0.42& 3.7  & 0.97402&    &       &    &         & 2  & 	      &		   \\
 15& 1.15& 10.1 & 0.98285&    &       &    &         & 2  &	      &		   \\
 16& 0.22& 3.7  & 0.98464&    &       &    &         & 2,3& 	      &		   \\
 17& 0.89& 14.9 & 1.03200&   9&  1.44 & 3.3&  1.03232& 3  &-47 \pm 45 &-0.55\pm0.44\\
 18& 0.23& 4.3  & 1.04635&    &       &    &         & 2  &	      &		   \\
 20& 0.27& 3.7  & 1.06408&    &       &    &         & 2  &	      &		   \\
 21& 0.78& 6.3  & 1.06791&    &       &    &         & 2,3&	      &		   \\
 22& 0.23& 5.5  & 1.08644&    &       &    &         & 1  &	      &		   \\
 23& 0.41& 4.7  & 1.10286&    &       &    &         & 1,3&	      &		   \\
 24& 0.49& 6.9  & 1.10726&    &       &    &         & 1  &	      &		   \\
 26& 0.30& 5.8  & 1.13146&    &       &    &         & 5  &  	      &		   \\
   & < 0.14&    &        &  20&  0.47 & 4.0&  1.14932&    &	      &		   \\
 28& 0.41& 9.1  & 1.16062&  21&  0.47 & 5.0&  1.16045&    &  23 \pm 26&-0.06\pm0.10\\
 36& 0.16& 3.6  & 1.24371&    &       &    &         & 5  &	      &		   \\
   & < 0.11&    &        &  29&  0.98 &12.3&  1.25970&    &	      &		   \\
 37& 0.39& 5.8  & 1.26119&    & < 0.19&    &	     &    &	      &		   \\
 38& 0.13& 3.8  & 1.29600&    &       &    &         & 1  &	      &		   \\
 41& 0.32& 4.4  & 1.30778&    &       &    &         & 1  &	      &		   \\
 42& 0.76& 20.0 & 1.32543&  33&  1.16 &10.0&  1.32556&    & -16 \pm 11&-0.39\pm0.12\\
 43& 0.18& 4.0  & 1.34335&  34&  0.30 & 3.0&  1.34331&    &   5 \pm 35&-0.12\pm0.11\\
 44& 0.19& 3.3  & 1.34589&    &       &    &         & 5  &	      &		   \\
   & < 0.10&    &        &  35&  0.26 & 3.6&  1.35237&    &	      &		   \\
   & < 0.10&    &        &  36&  0.44 & 3.2&  1.38781&    &	      &		   \\
 45& 0.27& 6.6  & 1.39100&  37&  0.61 & 4.5&  1.39094&    &   7\pm33  &-0.34\pm0.14\\
 46& 0.14& 3.1  & 1.39437&  38&  0.25 & 3.9&  1.39475&    & -47\pm32  &-0.10\pm0.08\\
 47& 0.97& 23.2 & 1.40817&    &       &    &         & 1  &	      &		   \\
 50& 0.46& 15.9 & 1.43016&  46&  0.23 & 6.2&  1.43075&    & -72 \pm 9 & 0.23\pm0.05\\
 51& 0.16& 5.5  & 1.43427&    &       &    &         & 2  &	      &		   \\
 52& 0.43& 15.7 & 1.44457&  49&  0.21 & 4.0&  1.44454&    &   4 \pm 27& 0.21\pm0.06\\
   & < 0.11&    &        &  50&  0.18 & 5.6&  1.44895&    &	      &		   \\
 53& 0.10& 3.7  & 1.45182&  51&  0.68 & 9.8&  1.45224& 6  &-51 \pm 22 &-0.58\pm0.03\\
 54& 0.14& 3.1  & 1.45373&  52&  0.52 & 9.1&  1.45444& 6  &-87 \pm 20 &-0.38\pm0.07\\
 55& 0.34& 7.5  & 1.45579&    &       &    &         & 6  &	      &		   \\
            \noalign{\smallskip}
            \hline
         \end{array}
      \]
\begin{list}{}{}
\item[1] Three sigma detection limits in B $>W$(A).
\item[2] Spectral region covered by the BAL troughs in B.
\item[3] \ion{C}{iv} associated.
\item[4] Maybe \ion{S}{iv} $\lambda$944 BAL.
\item[5] Absorption feature present in B at low significance.
\item[6] Associated system.
\end{list}

   \end{table*}


\subsection{Definition of Line Samples}
\label{1216_samples}

To define the number of coincidences and anti-coincidences we have
selected from all 36 Ly$\alpha$ lines observed in A (the spectrum with
better signal-to-noise) those ones (1) at $z\le1.43$; (2) not associated
with metal lines; (3) at wavelengths not  
covered by the BAL troughs in B; and (4) at wavelengths where  $3
\sigma$ detection limits in B are lower than the measured equivalent
width of the line in A. Lines in B that had ${\rm SNR}>3$ and fulfilled
criteria (1) to (3) were also selected.  
In some cases, lines in A have a corresponding absorption feature at
the same wavelength in B, but at a too low significance level to
unambiguously designate the line pair as a coincidence. These lines in
A were excluded (A26, A36, A44). Notice that criterion (3)
automatically prevents 
comparing line equivalent widths for which the uncertainties
introduced by the placement of the B continuum are large.

The final sample of Ly$\alpha$ lines suitable for this study is
composed of $N=11$ redshift systems within $1.16<z<1.43$, a range not
much smaller than the one allowed by conditions (1) to (4). Out of this number,
$N_{\rm C}=6$ systems show Ly$\alpha$ lines in both spectra, and $N_{\rm A}=5$ 
lines are detected in only one spectrum. The rest-frame
equivalent widths range from 0.14 to 0.76 \AA{} in A, and from 0.23 to
1.16 \AA{} in B. This sample is hereafter
called ``full sample''. 

\begin{figure}[t]
      \vspace{0cm}
\hspace{0cm}\epsfig{figure=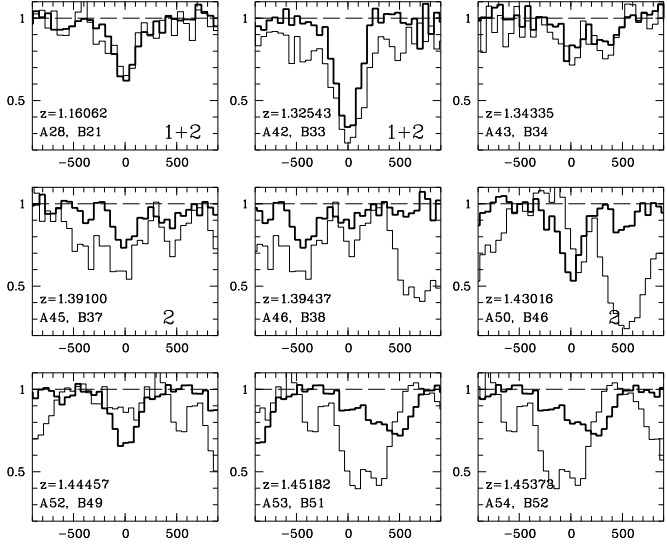,width=8.8cm}
\vspace{0cm}
      \caption[Ly$\alpha$ lines in
\hs]{Ly$\alpha$ lines present both in HS 1216+5032 A and 
      B, with ${\rm SNR}>3$, $\Delta v({\rm A-B})<100$ 
      \kms, and no metals associated. These lines lie at wavelengths
not covered by the BAL troughs in B. The right number indicates
      to which sample the line pair belongs. The three bottom panels show
      lines that fall within $3\,000$ \kms{} from the QSO
      redshift. The thick line 
      represents the flux of A. The abscissa is in \kms{} and the
      zero-velocity point is defined by the redshift of each A line. 
              }
         \label{fig3_1216}
   \end{figure}

Coincident lines are shown in
Figure~\ref{fig3_1216}, where the thick line 
represents the flux of A and the zero velocity point corresponds to the
redshift of the A line. Only the first six panels show lines in the 
``full sample''; the remaining lines arise in Ly$\alpha$ clouds likely
to be influenced by the QSO flux. All coincident lines are 
separated by less than 100 \kms. Anti-coincident lines are all uniquely
defined, as there are no corresponding lines in the other spectrum within
several hundred \kms. 

However, there are two exceptions that must be
pointed out. The first one is line 29 in B, separated by
$198$ \kms{} from the nearest Ly$\alpha$ line in A (37). These lines
in A and B have been counted as  anti-coincidences, though we are 
suspicious of this interpretation as the line profiles suggest line
B29 is blended with an absorption feature seen in both
spectra. The second is another of the 
anti-coincidences, line B50, which could alternatively be identified with
\ion{Mg}{i} at $z=0.04$. These cases will introduce unavoidable
uncertainties in the results presented here. Let us recall that the
exclusion of  
any one line from the samples would lead to significantly different results for
$R_{\rm c}$. This illustrates the 
real uncertainties dominating simulations with such a small number of
observed lines. Moreover, the samples are limited by the large redshift path
blocked out by the BAL troughs in B, so there must be a considerable loss
of information. 

   \begin{figure}[t]
      \vspace{0cm}
\hspace{0cm}\epsfig{figure=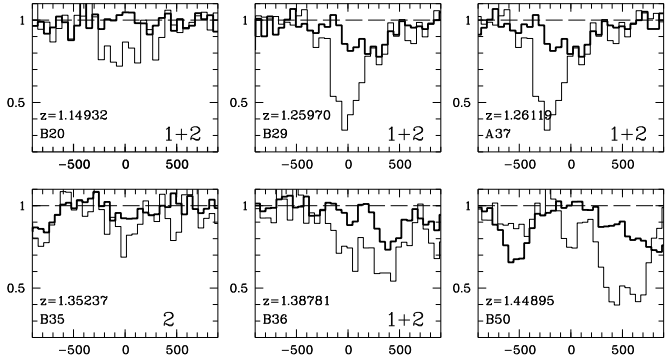,width=8.8cm}
\vspace{0cm}
      \caption[Ly$\alpha$ lines in
\hs]{Ly$\alpha$ lines present either in HS 1216+5032 A or 
      B, with ${\rm SNR}>3$, $\Delta v({\rm A-B})<100$ 
      \kms, and no metals associated. These lines lie at wavelengths
not covered by the BAL troughs in B. The right number indicates
      to which sample the line pair belongs. The abscissa is in \kms{} and the
      zero-velocity point is defined by the redshift of each present line. 
              }
         \label{fig3_1216_new}
   \end{figure}

A line significance level 
of $5$ for lines in A has been chosen to perform the 
likelihood analysis 
described in the next subsection . This selection
automatically excludes two of the redshift systems in the full sample
and defines the following sub-samples:
\begin{enumerate}
\item {\it Strong-line sample (sample 1):} lines for which SNR(A) $>5$,
SNR(B) $>3$, and $W_0>0.32$ \AA{}. This selection
yields $N_{\rm C}=2$ coincidences and $N_{\rm A}=4$ anti-coincidences.
\item {\it Strong+weak line sample (sample 2):} same as previous
sample but with 
$W_0>0.17$ \AA{}. This selection yields $N_{\rm C}=4$ and $N_{\rm A}=5$.
\end{enumerate}

\subsection{Maximum Likelihood Analysis}
\label{1216_max}

In the following we discuss a likelihood analysis on cloud sizes in
front of \hs. 
The technique is based on the
definition of a likelihood function $\cal{L}$ that gives the probability of
getting the observed number of coincidences and
anti-coincidences. Evidently, such a function must depend on the shape
of the absorber. Here we concentrate on two possible cloud
geometries for which $\cal{L}$ can be derived analytically: spheres
and cylinders. 

For coherent spherical clouds, the probability that one cloud at
redshift $z$ is
intersected by one LOS is given by McGill (\cite{McGill}) as:
\begin{equation}
\phi(X)=\frac{2}{\pi}\left\{ \arccos\left[ X(z)\right]
-X(z)\sqrt{1-X(z)^2}\right\} ~,
\end{equation}
where $X(z)\equiv S(z)/2R_{\rm c}$ and $R_{\rm c}$ is the cloud
radius. The samples consider only redshifts where at least one
LOS intersects a cloud  
(otherwise one should make assumptions on the cloud distribution along
the LOS). Therefore, one must compute the probability that, given that
a line appears in one spectrum, a line will appear in the other
spectrum. This probability is given by Dinshaw et
al. (\cite{Dinshaw2}) as: 
\begin{equation}
\psi(X)=\frac{\phi}{2-\phi} ~.
\end{equation}
Finally, the probability of getting the observed number of
coincidences and anti-coincidences is given by the product
\begin{equation}
{\cal L}(R_{\rm c})=\prod_i \psi\left[ X(z_i)\right] ~\prod_j 
\left\{ 1-\psi\left[ X(z_j)\right]\right\} ~,
\end{equation}
where the indexes $i$ and $j$ number the coincidences and
anti-coincidences, respectively. Note that if there is at least one
anti-coincidence, then $\cal L$ has a maximum value. Otherwise it
grows monotonically.

\begin{figure}[t]
      \vspace{0cm}
\hspace{0cm}\epsfig{figure=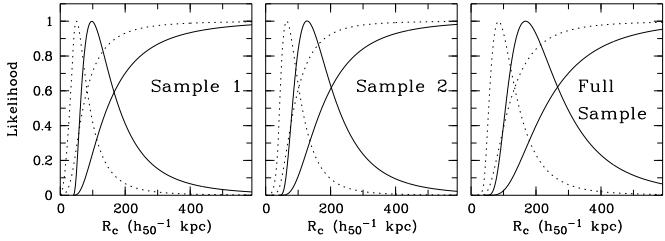,height=8.8cm, angle=-90}
\vspace{0cm}
      \caption[]{
      Likelihood function ${\cal L}(R_{\rm
      c})$ normalized  
      to its peak intensity vs. cloud  radius $R_c$, and cumulative
      distribution for spherical (solid line) and cylindric absorbers. 
              }
         \label{fig25}
   \end{figure}

The solid curve in Fig.~\ref{fig25} shows the results of the
likelihood function ${\cal L}(R_c)$ normalized to its peak intensity
for the various samples. 
The figure also shows the cumulative distribution of ${\cal
L}(R_c)$. The estimated $2 \sigma$ limits on cloud diameters for
\hs{} -- derived from the cumulative distribution -- are listed in 
Table~\ref{tbl-14}. 
The peak intensity of ${\cal L}$ is used to find the most
probably radii, i.e., $96$ \hkpc{}  for the strong line sample
and $128$ \hkpc{} for the strong+weak line sample. From the cumulative 
distribution, the respective median values are $145$ and $173$
\hkpc{}. In both cases, the derived sizes are larger if weaker
lines are used. This result
provides evidence that Ly$\alpha$ 
clouds must have a smooth density distribution. 


In a second model, let us suppose that Ly$\alpha$ lines occur in
filamentary structures lying perpendicular to the LOSs. Such
structures can
%
be idealized as cylinders with a
radius-to-length ratio $\ll 1$. 
For cylindric clouds, the probability that one LOS
intersects a cloud at
redshift $z$ given that the other LOS already does, reads (A. Smette,
private communication): 
\begin{equation}
\phi(X)=\left\{ 
  \begin{array}{ll}   
  \frac{2}{\pi} \bigg\{ \arcsin \left [ X(z)^{-1} \right ] &\\
  \hspace{1.em} + \sqrt{X(z)^{2} - 1} - X(z) \bigg\} & {\rm for} X(z)>1;\\
  &\\
  1-\frac{2}{\pi}~X(z) & {\rm otherwise},
  \end{array}
  \right. 
\end{equation}
where $X(z)\equiv S(z)/2R_{\rm c}$ and $R_{\rm 
c}$ is the  radius of the cylinder. 
The dotted curve in Fig.~\ref{fig25} shows the results of the
likelihood function ${\cal L}(R_c)$  normalized to its peak intensity
and the cumulative distribution for the various samples.
Most probably cylinder radii are  $49$, $65$, and $85$ \hkpc{}  for the strong,
strong+weak and full line samples, respectively.
The respective median values are $70$, $87$ and $115$
\hkpc{}. Two sigma bounds are displayed in Table~\ref{tbl-14}. 
Again, {\it the derived sizes are larger if weaker
lines are used}, providing evidence that the density distribution in
Ly$\alpha$  clouds must be smooth (see, e.g., Monier et
al.~\cite{Monier}). 
These size estimates are almost $50$\% lower than for spherical
absorbers. This can be explained
by the fact that, in general, elongated 
structures can more easily reproduce the observed number of
coincidences for a given radius than a sphere; in fact, the
probability $\phi(R_c)$ for spherical absorbers vanishes for $R_c<S/2$,
while for cylinders $\phi>0$ for all nonzero values of $R_c$. 
The lengths of such structures are much larger than these
values, but undefined in the model. 

   \begin{table}[t]
      \caption[Diameter of Ly$\alpha$ clouds in \hs]{ 
   Diameter $D$ of spherical and cylindric Ly$\alpha$ clouds in \hs{} in the
   redshift interval $z=1.15-1.43$ as derived from the likelihood
      function defined in (3). Values are in kpc and represent $2
   \sigma$ limits. $H_0=50$ \kms~Mpc$^{-1}$ and
   $q_0=0.5$.
}
         \label{tbl-14}
      \[
         \begin{array}{cccc}
            \hline
            \noalign{\smallskip}
{\rm Cloud~shape}&{\rm Sample 1}&{\rm Sample 2}&{\rm Sample 3}
\\ 
            \noalign{\smallskip}
            \hline
            \noalign{\smallskip}
{\rm Spherical}&136<D<880&172<D<896&224<D<1160\\  
            \noalign{\smallskip}
{\rm Cylindric}&52<D<454&86<D<460&112<D<610\\  
            \noalign{\smallskip}
            \hline
         \end{array}
      \]
\begin{list}{}{}
\item[] {Sample 1: Lines with $W_0>0.32$ \AA{}; 2 coincidences, 4
anti-coincidences.}
\item[] {Sample 2: Lines with $W_0>0.17$ \AA{}; 4 coincidences, 5
anti-coincidences.}
\item[] {Sample 3: Full sample: 6 coincidences, 5 anti-coincidences.}
\end{list}
   \end{table}

\subsection{Equivalent Widths}
\label{1216_ew}

\begin{figure}[t]
   \centering
      \vspace{0cm}
\hspace{0cm}\epsfig{figure=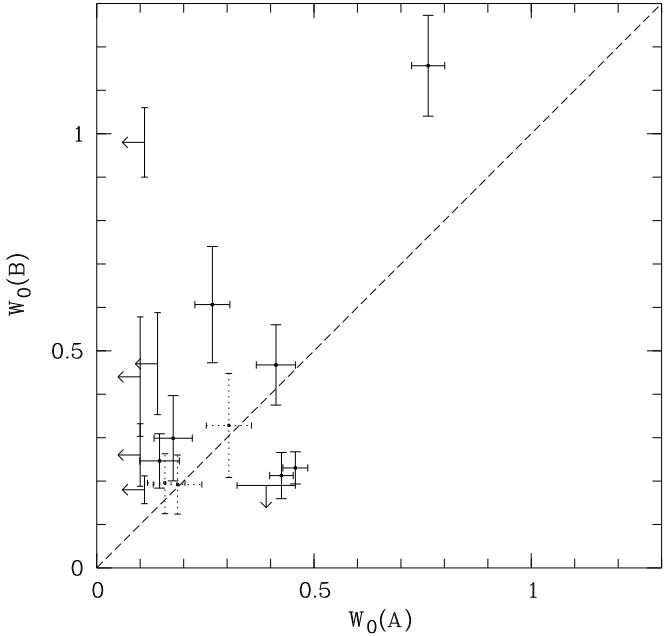,width=8.8cm}
\vspace{0cm}
      \caption[Equivalent widths of Ly$\alpha$ lines in \hs]{
Rest-frame equivalent widths of Ly$\alpha$ lines common to both
spectra, with ${\rm SNR}>3$, $\Delta v<100$ \kms,
and no metal lines associated. Dotted crosses represent line pairs
with $2.58<{\rm SNR}<3$. The ``associated system'' has not been
      considered. Lines detected in only one spectrum are also
      shown, with upper limits representing $3 \sigma$ detection
      limits in the other spectrum.
              }
         \label{fig4_1216}
   \end{figure}

The sample of Ly$\alpha$ lines common to both spectra gives
model-independent information about the size scales of the absorbers. 
Figure~\ref{fig4_1216} shows the rest-frame
equivalent widths of lines  
common to both spectra and their $1 \sigma$ errors, where the dashed
straight line has a slope of unity.  Only lines without
associated metal lines are included. Note that this selection yields
line-pairs with $\lambda>2620$ \AA{}, so the influence of uncertainties in the
continuum placement (e.g. due to the BAL troughs or to spectral
regions with high noise level) over the equivalent-width estimates is
minimized. We have included  
one line pair within 3\,000 \kms{} from the QSO's redshift
(A52,B49) but not the ``associated system''. The dotted crosses
represent line pairs for which  the B line has  $2.58<{\rm SNR}<3$,
i.e., those ones labeled  with footnote 5 in Table~\ref{tbl-4_1216}. 
Additionally, to illustrate the significance of the non-detections we
have also included lines detected in only one spectrum, plotted
against the corresponding $3 \sigma$ detection limit in the other
spectrum.  

Most of the coincident lines are
stronger in the B spectrum, which is explained by the difficulty in
detecting lines in B (an additional consequence of this is that 
most anti-coincidences are lines in B). It can be seen that there are some
significant deviations from $\Delta W_0=0$, even excluding the 
line pair within 3\,000 \kms. For the latter such deviation
would be expected, since the QSOs themselves might be an important ionizing
agent in clouds next to the QSOs.\footnote{Indeed, if the QSO
radiation field contributed significantly to ionize
these clouds, one would expect $W({\rm A})<W({\rm B})$ for
these lines, because QSO A is more luminous than B. Such difference in
equivalent width is, however, not observed.}  

\begin{figure}[t]
    \centering
      \vspace{0cm}
\hspace{0cm}\epsfig{figure=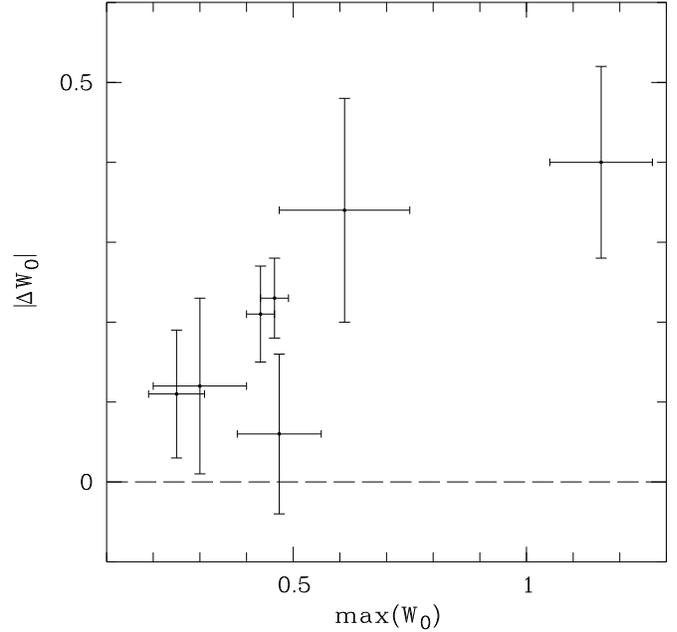,width=8.8cm}
\vspace{0cm}
      \caption[\hs: Equivalent width differences vs. $\max(W_0({\rm
A}),W_0({\rm B}))$]{ 
      Absolute value of rest-frame equivalent width differences (\AA) 
      vs. $\max(W_0({\rm A}),W_0({\rm B}))$ of Ly$\alpha$ lines with
      ${\rm SNR}>3$, $\Delta v<100$ \kms, and no metal lines
      associated.  The ``associated system'' has not been
      considered.  
              }
         \label{fig26}
   \end{figure}

The equivalent width differences $\Delta W_0 =
|W_0({\rm A})-W_0({\rm B})|$ show no
correlation with $\Delta v$. Instead, $\Delta W_0$ and max$(W_0({\rm
A}),W_0({\rm B}))$ seem to be correlated (see Fig.~\ref{fig26}), with
larger equivalent width differences for  
larger equivalent widths (e.g. Fang et al.~\cite{Fang}). Both effects,
although not statistically significant, 
suggest coherent structures, i.e., no ``cloudlets'' at similar
redshifts (Charlton et al.~\cite{Charlton}). In conclusion, LOS A
and B must sample coherent clouds showing small -- but 
otherwise significant -- density gradients on spatial scales of $\sim
80~h_{50}^{-1}$kpc, the linear separation between LOSs in this redshift
range. 

\subsection{The $z_a\approx z_e$ Ly$\alpha$ systems in HS 1216+5032 A and B}
\label{1216_assoc}

Two $z_a\approx z_e$ Ly$\alpha$ systems are observed in the
spectra of both  HS 1216+5032 A and B, through the partially
resolved Ly$\alpha$ lines A53, A54, A55, and B51 and B52 
(see Fig.~\ref{fig3_1216}). The lines in A
and B differ by $\Delta v_{\rm A53-B51}=-51$ \kms and $\Delta v_{\rm
A54-B52}=-87$ \kms. The system in A is blueshifted by 122 \kms{}
relative to $z_e({\rm A})$, while the system in B is redshifted by 280
\kms{} relative to $z_e({\rm B})$ (but see
\S~\ref{1216_em}). Line A55 has no clear counterpart in B. 
No metal lines are found associated with these systems. The total 
equivalent width of the A lines is larger than in B
by a factor of 2.1. 

The absence of highly ionized species such as
\ion{N}{v} and \ion{O}{vi} 
suggests that these systems  are probably not physically
associated with the 
QSOs (e.g., Turnshek~\cite{Turnshek1};
Petitjean~\cite{Petitjean1}). Furthermore, the small velocity  
differences between lines in A and B, and the 
fact that both systems have two line components well correlated in
redshift, strongly suggest they arise in the same absorber
(see also Petitjean et al.~\cite{Petitjean}; Shaver \&
Robertson~\cite{Shaver}). If this is true, 
the different Ly$\alpha$ line strengths between A and B imply
either characteristic transverse sizes of $\sim 80~h_{50}^{-1}$kpc 
or maybe even larger clouds with ionization conditions that change on
such scales. The size scales are compatible with the idea that these systems
arise, for example, in the very extended halo of an intervening galaxy
(Weymann et al.~\cite{Weymann}), or in the intra-group gas of the QSO 
host galaxies. 

If the radiation field of the QSOs is the main ionization source
in the clouds giving rise to these systems (assuming the QSO pair is
real), then the systems in B should be less ionized than the A
ones. This is because QSO A is intrinsically more luminous than
B in the ionizing continuum.
Consequently, the
different line 
strengths in A and B can also be explained if the LOSs cross
regions of similar density, with LOS B probing regions with more
neutral gas. Of course, the latter case does not rule out the
intervening nature of these systems.

\subsection{Discussion}
\label{lya_conclusions}

Fang et al. (\cite{Fang}) have pointed out that there seems to be a trend
of larger estimated cloud sizes with increasing LOS separation
$S$ (see their Fig. 5). If such a trend were real, it would
imply that the scenario of uniform-sized spherical clouds is too
idealized. Their data, however, are insufficient to discern
whether the effect is due to non-uniform cloud size or simply
non-spherical cloud geometry. The results presented in our work 
(slightly modified by defining a sample of lines with $W_0>0.40$
\AA{} to be consistent with these authors) fit this tendency  well because
they add a new measurement of relatively small
cloud sizes at small LOS separation. We note, however, that 
D'Odorico et al. (\cite{D'Odorico}), who use a larger database and an
improved statistical approach, find no correlation between cloud size
and LOS separation. Clearly, more QSO pairs are needed that span a
range in LOS separations to confirm or discard the size/separation
correlation. 

Here we propose that the notion of a single population of uniform-sized clouds
must be revised. In fact, if Ly$\alpha$ clouds span a range of sizes between,
say, a few hundred kpc to a few Mpc, then LOSs to QSO pairs with arcminute
separations would be crossing not only huge and coherent
structures, but also smaller clouds correlated in redshift. 
Further evidence for a more complicated
scenario than usually assumed comes from cosmological hydrodynamic
simulations made for $z=3$, which show 
entities of {\it non-uniform} sizes grouped along filamentary structures (e.g.,
Cen \& Simcoe~\cite{Cen}). However, we recall that the situation at
lower redshift can be different if such structures do evolve in size.

Size evolution of Ly$\alpha$ absorbers has not yet been 
observed, in part because the number of adjacent QSOs with suitable
angular separations is not statistically significant, but also due to
the scarce number of observations at low redshift. In particular, the
present data on \hs{} ($<z>=1.3$) do {\it not} confirm the suggestion (Fang et
al.~\cite{Fang}; Dinshaw et al.~\cite{Dinshaw1}) 
that cloud sizes increase with decreasing redshift. This result is in
variance with the findings by D'Odorico et al. (\cite{D'Odorico}).

Assuming that the Ly$\alpha$ absorbers are filamentary
structures -- modelled as cylinders of infinite length -- lying
perpendicular to 
the LOSs, our likelihood analysis leads to almost $50 \%$
smaller transverse dimensions than 
for spherical clouds. Unfortunately, the 
method presented here is not capable of distinguishing between cylinders and
spheres. Moreover, the information provided by two LOSs might 
be insufficient to make such a distinction possible, so one should use
triply imaged QSOs (Crotts \& Fang~\cite{Crotts1}) or, even better,
QSO groups (Monier et al.~\cite{Monier}). 
It is worth saying, however, that flattened structures are
more able to simultaneously reproduce the 
requirements of neutral gas density from photoionization models and
the transverse scale lengths derived from double LOSs than
spherical absorbers do. Such photoionization models lead to structures
with thickness-to-length ratios of $\sim 1/30$ (Rauch \& Haehnelt
\cite{Rauch4}). Indeed, hydrodynamical
simulations in the context of hierarchical structure formation 
have shown that at $z\sim3$, high density gas regions (producing
$W_0>0.3$ \AA{} Ly$\alpha$ absorption lines) are connected by
filamentary and sheet-like structures roughly $10^{6}$ times less dense
than the embedded condensations. The filaments seem to evolve slowly
and still fill the Universe at $z\sim1$ (Dav\'e et 
al.~\cite{Dave}; Riediger, Petitjean \& M\"ucket~\cite{Riediger}),
giving rise to the majority of strong Ly$\alpha$ lines.
If this picture is correct, the requirement of different 
cloud populations might not be necessary to explain current
observations. 

New high-resolution ultraviolet observations of QSO pairs are
needed. Despite technical difficulties (close QSO pairs tend to have
such different observed fluxes that high dispersion spectroscopy
normally leads to at  least one spectrum with high noise levels) 
they should 
improve considerably our knowledge of the Ly$\alpha$-cloud geometry by 
(1) analyzing absorbers that produce only weak ($W_0\sim0.01$ \AA)
lines, i.e., those surely not associated with low-brightness galaxies,
and  (2) testing models that consider column density distributions. 

\section{Summary}
\label{sec_dis}

We have presented {\it HST} FOS spectra of the QSO pair HS
1216+5032 AB. Our results are summarized as follows:
\begin{enumerate}
\item{{\em Metal systems.} 
Three strong and complex \ion{C}{iv} absorption systems are observed
in both {\it HST} spectra at $z\approx0.72$. The velocity span along
the LOSs, $\Delta 
v\sim1500$ \kms, 
is large and suggests 
the gas where this \ion{C}{iv} occurs might be associated
with a cluster of galaxies. Differences in
the line strengths suggest that, if these systems arise in
coherent structures, they must have characteristic transverse sizes of
$\sim75~h_{50}^{-1}$kpc. Alternatively, the systems may be composed of a
large number of cloudlets correlated in redshift. 

In the spectrum of B, a strong
absorption feature at $\lambda\sim 2910$ \AA{} is identified with two
\ion{Mg}{ii} doublets at $z=0.04$. If this identification is correct,
these lines could arise in a low-redshift damped Ly$\alpha$ system. 
}
\item{{\it The BAL systems in B:}
The spectrum of QSO image B shows BAL troughs by \ion{H}{i}, \ion{C}{ii},
\ion{C}{iii}, \ion{N}{iii}, \ion{N}{v}, \ion{O}{vi}, and possibly
\ion{S}{iv} and \ion{S}{vi} arising in absorption
systems at outflow velocities from the QSO of up to $5000$
\kms. The BAL troughs arise probably as a consequence of absorption by a
mixture of broad and narrow components. 
}
\item{{\em Ly$\alpha$ absorbers.} 
Due to the redshift path blocked out by
the BAL troughs in B the number of detected lines is small. Selecting
lines not associated with 
metal lines, with $\Delta v({\rm A-B})<100$ \kms{} and $W_0>0.17$ \AA{}
yields four lines common to both spectra and five lines without
counterpart in the other spectrum. For $W_0>0.32$ \AA{} lines the 
numbers are two and four, respectively. Using a maximum likelihood
technique, most probably diameters for spherical clouds of $192$ 
and $256$ \hkpc{} are found for $W>0.32$ \AA{} and $W>0.17$ \AA{}
lines, respectively. The $2 \sigma$ limits derived using the 
cumulative distribution of the probability function are
$136<D<880$ \hkpc{} and $172<D<896$ \hkpc{} for the respective samples
at $<z>=1.3$. 
Assuming that the absorbers are filamentary structures
lying perpendicular to the LOSs, transverse dimensions almost $50 \%$
smaller than for spherical clouds are found. In both cases, the 
results of our analysis do not confirm the claim that the
characteristic size of the Ly$\alpha$ absorbers increases with
decreasing redshift. 

Independently of the cloud models used, we note
that there are
significant equivalent width differences 
between lines in A and B. Also, there appears to be a trend of larger
equivalent width differences 
with increasing line strength, while no velocity differences between
common lines is found. This provides evidence that the absorbers are
coherent entities. The results for each line sample 
suggest that the absorbers must
have a smooth gas density distribution, with lower density gas being
more extended. 
}

\end{enumerate}

\begin{acknowledgements}
We are grateful to Michelle Mizuno, Lutz Wisotzki, and specially Alain
Smette for their valuable comments on  early drafts. We also thank
Olaf Wucknitz for calculations concerning the gravitational-lens hypothesis. 
S. L. acknowledges support by the BMBF (DARA) under grant No. 50 OR 9905.
\end{acknowledgements}


\begin{thebibliography}{}
    \bibitem[1999a]{Arav1}
      Arav, N., Becker, R. H., Laurent-Muehleisen S. A., Gregg, M. D.,
      White, R. L., \& de Kool, M. 1999a, ApJ 524, 566
    \bibitem[1999b]{Arav}
      Arav, N., Korista K. T., de Kool M., Junkkarinen V. T.,
      \& Begelman M. C. 1999b, ApJ 516, 27
    \bibitem[1997]{Barlow}
      Barlow, T. A., \& Sargent W. L. W., 1997, AJ 113, 136 
    \bibitem[1994]{Bechtold}
      Bechtold, J,, Crotts, A. P. S., Duncan, R. C., \& Fang, Y, 1994,
      ApJ 437, L83  
    \bibitem[1986]{Bergeron2} 
      Bergeron, J., \& Stasinska, G., 1986, A\&A 169, 1
    \bibitem[1991]{Bergeron} 
      Bergeron, J., \& Boiss\'e, P., 1991, A\&A 243, 344
    \bibitem[1997]{Cen}
      Cen, R., \& Simcoe, R. A., 1997, ApJ 483, 8 
    \bibitem[1995]{Charlton}
      Charlton, J. C., Churchill, C. W., \& Linder S. M., 1995, ApJ
      452, L81 
    \bibitem[1994]{Crotts}
      Crotts, A. P. S., Bechtold, J., Fang, Y., \& Duncan R. C., 1994, ApJ
      437, L79  
    \bibitem[1998]{Crotts1}
      Crotts, A. P. S., \& Fang, Y. 1998, ApJ 502, 16  
    \bibitem[1999]{Dave}
      Dav\'e, R., Hernquist, L., Katz, N.,  \& Weinberg, D. H., 1999,
      ApJ 511, 521   
    \bibitem[1994]{Dinshaw}
      Dinshaw, N., Impey, C. D., Foltz, C. B., Weymann, R., \& Chaffee,
      F. H., 1994, ApJ 437, L87 
    \bibitem[1998]{Dinshaw1}
      Dinshaw, N., Foltz, C. B., Impey, C. D., \& Weymann, R. J., 1998,
      ApJ 494, 567 
    \bibitem[1997]{Dinshaw2}
      Dinshaw, N., Weymann, R. J., Impey, C. D., Foltz, C. B., Morris,
      S. L., \& Ake, T., 1997, ApJ 491, 45 
    \bibitem[1998]{D'Odorico}
      D'Odorico, V., Cristiani, S., D'Odorico, S., Fontana, A., Giallongo, E.,
      \& Shaver, P., 1998, A\&A 339, 678
    \bibitem[1996]{Fang}
      Fang, Y., Duncan, R. C., Crotts, A. P.,\& Bechtold J, 1996, ApJ
      462, 77 
     \bibitem[1995]{Hagen1} 
      Hagen, H.-J., Groote, D., Engels, D., \& Reimers, D.,
      1995, A\&AS 111, 195
     \bibitem[1996]{Hagen} 
      Hagen, H.-J., Hopp, U., Engels, D., \& Reimers, D.,
      1996,  A\&A 308, L25
    \bibitem[1995]{Hamman1} 
      Hamann, F., Barlow, T. A., Beaver, E. A., Burbidge, E. M.,
      Cohen, R. D., Junkkarinen, V., \& Lyons, R.
      1995,  ApJ 443, 606
    \bibitem[1997]{Hamman} 
      Hamann, F., Barlow, T. A., Junkkarinen, V., \& Burbidge, E. M.
      1997,  ApJ 478, 80
    \bibitem[1999]{Kochanek} 
      Kochanek, C. S., Falco, E. E., \& Mu\~noz, J. A., 1999,  ApJ 510, 590
    \bibitem[1995]{Kochanek1} 
      Kochanek, C. S. 1995,  ApJ 453, 545
    \bibitem[1996]{Lu}
      Lu, L. Sargent, W. L. W., Barlow, T. A., Churchill, C. W., \& Vogt,
      S. S., 1996, ApJS 107, 475
    \bibitem[1990]{McGill} McGill, C. 1990, MNRAS, 242, 544 
    \bibitem[1999]{Monier}
      Monier, E. M., Turnshek, D. A., Hazard, C., 1999, ApJ 522, 627 
    \bibitem[1994]{Petitjean1}
      Petitjean, P., Rauch, M., Carswell, R. F., 1994, A\&A 291, 29
    \bibitem[1998]{Petitjean}
      Petitjean, P., Surdej, J., Smette, A., Shaver, P., M\"ucket, J.,
      \& Remy, M., 1998, A\&A 334, L45
    \bibitem[1995]{Rauch4} 
      Rauch, M., \& Haehnelt, M. G. 1995, MNRAS, 275, 76
      ApJ, 467, L5  
    \bibitem[1997]{Rauch1} 
      Rauch, M. 1997, in {\it Structure and Evolution of the IGM from
      QSO Absorption Lines}, Proc. 13th IAP Colloquium,
      ed. P. Petitjean, S. Charlot (Paris: Editions Fronti\`eres), 109 
    \bibitem[1998]{Riediger} 
      Riediger, R., Petitjean, P., \& M\"ucket, J. P. 1998, A\&A 329,
      30   
    \bibitem[1989]{Sargent} 
      Sargent, W. L. W., Steidel, C. C., \& Boksenberg, A. 1989, ApJS
      69, 703 
    \bibitem[1993]{Savage} 
      Savage, B. D., Lu, L., Bahcall, J. N., Bergeron, J., et al.
      1996, ApJ 413, 116
    \bibitem[1993]{Schneider} 
      Schneider, D. P. et al. 1993, ApJS 87, 45
    \bibitem[1983]{Shaver} 
      Shaver, P. A., \& Robertson J. G. 1992, ApJ 268, L57
    \bibitem[1992]{Smette1} 
      Smette, A., Surdej, J., Shaver, P. A., Foltz, C. B., Chaffee,
      F. H., Weymann, R. J., Williams, R. E., \& Magain, P. 1992,
      ApJ 389, 39
    \bibitem[1995]{Smette} 
      Smette, A., Robertson, J. G., Shaver, P. A., Reimers, D., Wisotzki,
      L., \& K\"ohler, Th. 1995, A\&AS 113, 199
    \bibitem[1992]{Steidel} 
      Steidel, C. C., \& Sargent W. L. W.  1992, ApJS 80, 1
    \bibitem[1984]{Turnshek1} 
      Turnshek, D. A. 1984, ApJ 280, 51
    \bibitem[1996]{Turnshek} 
      Turnshek, D. A., Kopko, M., Monier, E., \& Noll, D. 1996, ApJ
      463, 110
    \bibitem[1992]{Tytler} 
      Tytler, D., \& Fan, X.-M. 1992, ApJS 79, 1
    \bibitem[1994]{Verner}
      Verner, D. A., Barthel, P. D., \& Tytler, D. 1994, A\&AS 108, 287 
    \bibitem[1995]{Wampler}
      Wampler, E. J., Chugal, N. N., \& Petitjean P. 1995,
      ApJ 443, 586
    \bibitem[1979]{Weymann}
      Weymann, R. J., Williams, R. E., Peterson, B. M., \& Turnshek,
      D. A. 1979, ApJ 234, 33
    \bibitem[1991]{Weymann2}
      Weymann, R. J., Morris, S. L., Foltz, C. B., \& Hewett, P. C.
      1991, ApJ 373, 23
    \bibitem[1998]{Weymann1}
      Weymann, R. J., Jannuzi, B. T., Lu, L., et al. 1998, ApJ 506, 1 



\end{thebibliography}
\end{document}